\documentclass{article}

\usepackage{authblk}
\usepackage[square,numbers]{natbib}
\usepackage{mathtools, amsmath, amsthm, amssymb, amsbsy, amsfonts, amscd}
\usepackage{bm}
\usepackage{breqn}
\usepackage{color, soul}
\usepackage{enumerate}
\usepackage{empheq}
\usepackage{rotating}
\usepackage{ftnxtra}
\usepackage{fnpos}
\usepackage[titletoc,toc,title]{appendix}
\usepackage{euscript}
\usepackage{graphicx}
\usepackage{epsfig}
\usepackage{epstopdf}
\DeclareGraphicsExtensions{.pdf,.png,.jpg,.eps}
\usepackage{pstool}

\usepackage{psfrag}

\numberwithin{equation}{section}

\theoremstyle{plain}	
\newtheorem{thm}{Theorem}[section]

\theoremstyle{definition}	

\newtheorem{remark}[thm]{Remark}
\newtheorem{example}[thm]{Example}

\usepackage{caption}
\usepackage{subcaption}

\usepackage{hyperref}
\hypersetup{colorlinks=true, linkcolor=blue}
\hypersetup{colorlinks=true,citecolor=blue}

\DeclareMathAlphabet{\mathpzc}{OT1}{pzc}{m}{it}

\usepackage{amsmath, amsthm, amssymb}

\usepackage{cleveref}

\begin{document}
\bibliographystyle{abbrvnat}

\title{\textbf{Line and Point Defects in Nonlinear Anisotropic Solids}}

\author[1]{Ashkan Golgoon}
\author[1,2]{Arash Yavari\thanks{Corresponding author, e-mail: arash.yavari@ce.gatech.edu}}
\affil[1]{\small \textit{School of Civil and Environmental Engineering, Georgia Institute of Technology, Atlanta, GA 30332, USA}}
\affil[2]{\small \textit{The George W. Woodruff School of Mechanical Engineering, Georgia Institute of Technology, Atlanta, GA 30332, USA}}

\maketitle

\begin{abstract}
In this paper, we present some analytical solutions for the stress fields of nonlinear anisotropic solids with distributed line and point defects. In particular, we determine the stress fields of i) a parallel cylindrically-symmetric distribution of screw dislocations in infinite orthotropic and monoclinic media, ii) a cylindrically-symmetric distribution of parallel wedge disclinations in an infinite orthotropic medium, iii) a distribution of edge dislocations in an orthotropic medium, and iv) a spherically-symmetric distribution of point defects in a transversely isotropic spherical ball.
\end{abstract}

\begin{description}
	\item[Keywords:] Transversely isotropic solids; orthotropic solids; monoclinic solids; defects; disclinations; dislocations; nonlinear elasticity.
	\item[Mathematics Subject Classification] 74B20 $\cdot$ 70G45 $\cdot$ 74E10 $\cdot$ 15A72 $\cdot$ 74Fxx
\end{description}

\tableofcontents

\section{Introduction}
 \label{Intro}
In anelasticity, any measure of strain has both an elastic and a non-elastic part. Given a pair of thermodynamically conjugate stress and strain, locally a non-vanishing strain does not necessarily correspond to a non-vanishing stress. \emph{Elastic} strain refers to the part of strain that is locally related to the corresponding stress. The remaining part is referred to as \emph{eigenstrain}, a term that was first used by \citet{mura2012}. Defects are one source of anelasticity. Vito Volterra, in his seminal work \citep{volterra1907equilibre}, pioneered the mathematical study of defects many years before the first experimental observations of defects in solids. He classified line defects into six types, three of which are now called dislocations or translational defects, and the other three are called disclinations or rotational defects. 
\citet{kondo1955geometry,kondo1955non} and \citet{bilby1955continuous} independently explored the profound connections between the mechanics of defects and non-Riemannian geometries in the 1950s. \citet{kondo1955geometry,kondo1955non} discovered that the reference configuration of a solid is not necessarily Euclidean in the presence of defects. He realized that the curvature and the torsion of the reference manifold are measures of incompatibility and the density of dislocations, respectively. 
Defects due to plastic deformations naturally occur in most of the known problems in mechanics and tribology, e.g., contact mechanics \citep{jackson2005finite,brake2012analytical,jackson2013contact,brake2015analytical,jackson2015solution}, mechanical impact \citep{ghaednia2016permanent,kardel2017experimental}, and dislocation-boundary interactions \citep{wang2013pure,hooshmand2017atomistic}.
Other examples of anelastic sources include swelling and cavitation \citep{pence2005swelling,goriely2010elastic,moulton2011anticavitation}, bulk and surface growth \citep{amar2005growth,Yavari2010,Sozio201712}, thermal strains \citep{stojanovic1964finite,ozakin2010geometric,sadik2015geometric}, and the presence of inclusions and inhomogeneities \citep{yavari2013nonlinear,golgoon2016,golgoon2016torus,Golgoon2017}. There have been some theoretical investigations on the effects of eigenstrains in linear anisotropic media, e.g., \citep{willis1964anisotropic,li1998anisotropic,kinoshita1971,giordano2009}, and references therein.

Very little is known about the effects of material anisotropies on the stress field and energetics of defects in solids.
The dynamical response of uniformly moving dislocations in linear anisotropic media was studied by \citet{teutonico1962uniformly}. It was observed that both edge and screw dislocations are prone to exhibiting anomalous dynamical behavior such that the interaction force between two parallel dislocations (on the same slip plane) changes sign when dislocation velocity increases. \citet{head1967unstable} predicted instabilities of dislocations in some anisotropic metallic crystals. It was found that a straight dislocation may decrease its energy if it changes to a zig-zag shape, i.e., a straight dislocation may be unstable.
In the setting of the linear theory of elasticity, \citet{willis1970stress} analyzed dislocations in anisotropic media (see also \citep{willis1967second}). Particularly, the displacement fields of infinite straight dislocations and plane curvilinear dislocation loops were obtained. \citet{eshelby1949} investigated edge dislocations with an infinite straight axis in linear anisotropic solids. He extended Nabarro's calculation of the width of a dislocation to the anisotropic case. His results are limited to edge dislocations with an axis that is an infinite straight line, but there is no restriction on the type of anisotropy of the medium. 
\citet{schaefer1975elastic} investigated the elastic interaction of point defects in linear isotropic and anisotropic cubic media using Green's function approach. They specifically discussed the differences between the interactions in isotropic and anisotropic materials and the effects of anisotropy on the interaction potential. Some basic developments in the linear theory of dislocations in anisotropic media was given in \citep{lothe1992dislocations}. Methods for obtaining the induced linear elastic fields of defects in transversely isotropic bimaterials and orthotropic bicrystals (in 2D) were proposed in \citep{yu1995elastic} and \citep{yu2001two}, respectively. In particular, some closed-form solutions for inclusions and dislocation lines were presented.

A successive-approximation method was proposed in \citep{teodosiu1982} to study the nonlinear screw dislocation problem using the linear elasticity solution. Nonetheless, the method fails to find the correct solution near the dislocation axis. Only a handful of exact solutions for defects in nonlinear elastic solids exist in the literature, and they are all restricted to isotropic materials. We should mention \citep{Gairola,zubov1997,rosakis1988screw,ericksen1998non,acharya2001model,fedelich2004glide,YavariGoriely2012a,Sadik20160659} for dislocations, \citep{zubov1997,derezin2011discl,yavari2013riemann} for disclinations, and \citep{YavariGoriely2012c,yavari2014geometry,clayton2017finsler} for point defects and discombinations. 

To the best of our knowledge, despite the known importance of the anisotropic behavior of solids, especially at finite strains, the study of defects in the setting of nonlinear elasticity has been limited to isotropic solids. In this paper we study several examples of line and point defects in nonlinear anisotropic solids and present some analytical solutions for their stress fields. We consider an arbitrary cylindrically-symmetric distribution of parallel screw dislocations in orthotropic and monoclinic media, along with a parallel cylindrically-symmetric distribution of wedge disclinations in an infinite orthotropic medium. As the geometry of the material manifold explicitly depends on the distribution of defects, the material preferred directions (that identify the type of anisotropy) in the reference configuration explicitly depend on the defect distribution as well, and, in general, are different from those of the material in its current configuration. For instance, for the distributed  screw dislocations that we consider, the assumption that the dislocated body is orthotropic in the reference (current) configuration implies that the body is monoclinic in the current (reference) configuration. 

The boundedness of the stress components on the dislocation and disclination axes will be discussed. In particular, for an arbitrary cylindrically-symmetric distribution of parallel screw dislocations the stress exhibits a logarithmic singularity on the dislocation axis unless the axial deformation is suppressed. Note that these singularities arise due to the anisotropic effects (e.g., radial fiber-reinforcement), and, in particular, do not occur when the material is isotropic. Exploiting the so-called standard reinforcing model (see, e.g., \citep{merodio2003instabilities}), we obtain conditions under which the energy per unit length and the resultant longitudinal force of a single screw dislocation for a fiber-reinforced material are finite provided that the isotropic base material has a finite axial force and a finite energy per unit length. Employing Cartan's moving frames approach, for a given distribution of edge dislocations we will construct the material manifold and obtain explicit solutions for the stress field when the medium is orthotropic. We will also consider a spherically-symmetric distribution of point defects in a finite transversely isotropic spherical ball. We will show that for an arbitrary incompressible transversely isotropic material with the radial material preferred direction a uniform point defect distribution induces a uniform hydrostatic stress inside the region the distribution is supported.

The rest of the paper is structured as follows. In \S\ref{Prel} we tersely review some fundamentals of geometric nonlinear anisotropic elasticity and some related topics on nonlinear defect mechanics. We consider a cylindrically-symmetric distribution of parallel screw dislocations in orthotropic and monoclinic media in \S\ref{scr-or} and \S\ref{scr-mo}, respectively. A cylindrically-symmetric distribution of parallel wedge disclinations in an orthotropic medium is studied in \S\ref{wed-or}. In \S\ref{edge} edge dislocations in an orthotropic medium are considered. In \S\ref{eballc} we calculate the residual stresses due to a spherically-symmetric distribution of point defects in a transversely isotropic ball. We end the paper with some remarks in \S\ref{conc}.

\section{Geometric Anelasticity for Anisotropic Solids}
\label{Prel}

In this section we briefly review some fundamental elements of the geometric theory of nonlinear elasticity for anisotropic solids. For more detailed discussions, see \citep{MarsdenHughes1994,Yavari2006}.

\paragraph{Kinematics.}A body $\mathcal{B}$ is identified with a Riemannian manifold $\left(\mathcal{B},\mathbf G\right)$, and a configuration of $\mathcal{B}$ is a smooth embedding $\varphi:\mathcal{B}\rightarrow\mathcal{S}$, where $\left(\mathcal{S},\mathbf g\right)$ is a Riemannian manifold \textemdash the ambient space. An affine connection $\nabla$ on a smooth manifold $\mathcal{B}$ is a linear mapping~$\nabla:\mathcal{X}(\mathcal{B})\times\mathcal{X}(\mathcal{B})\rightarrow\mathcal{X}(\mathcal{B})$, where $\mathcal{X}(\mathcal{B})$ represents the set of all smooth vector fields on $\mathcal{B}$, such that the following properties are satisfied  $\forall~\mathbf{X},\mathbf{Y},\mathbf{X}_1,\mathbf{X}_2,\mathbf{Y}_1,\mathbf{Y}_2\in\mathcal{X}(\mathcal{B}),\forall~f,f_1,f_2\in C^{\infty}(\mathcal{B}),\forall~a_1,a_2\in\mathbb{R}$ (see~\citep{Docarmo1992,petersen2006riemannian} for more details): a) $\nabla_{f_1\mathbf{X}_1+f_2\mathbf{X}_2}\mathbf{Y}=f_1\nabla_{\mathbf{X}_1}\mathbf{Y}+f_2\nabla_{\mathbf{X}_2}\mathbf{Y}$, b) $\nabla_{\mathbf{X}}(a_1\mathbf{Y}_1+a_2\mathbf{Y}_2)=a_1\nabla_{\mathbf{X}}(\mathbf{Y}_1)+a_2\nabla_{\mathbf{X}}(\mathbf{Y}_2)$, c) $\nabla_{\mathbf{X}}(f\mathbf{Y})=f\nabla_{\mathbf{X}}\mathbf{Y}+(\mathbf{X}f)\mathbf{Y}$.
It can be shown that there is a unique torsion-free and compatible affine connection associated with any Riemannian manifold that is called a Riemannian connection. Let us denote the Levi-Civita connection associated with the Riemannian manifolds $\left(\mathcal{B},\mathbf{G}\right)$ and $\left(\mathcal{S},\mathbf g\right)$ by $\nabla^{\mathbf G}$ and $\nabla^{\mathbf g}$, respectively. We denote the set of all configurations of $\mathcal{B}$ by $\mathcal{C}$. A motion is a curve $c:\mathbb{R}^+\rightarrow\varphi_t\in\mathcal{C}$ such that $\varphi_t$ assigns a spatial point $x=\varphi_t(X)=\varphi\left(X,t\right)\in\mathcal{S}$ to every material point $X\in\mathcal{B}$ at any time $t$. The body is assumed to be   stress-free in its reference configuration, which may have a nontrivial geometry, in general, e.g., in the presence of eigenstrains. The deformation gradient $\mathbf{F}$ is the tangent map of $\varphi$ defined as $\mathbf{F}(X,t)=d\varphi_t(X): T_X\mathcal{B}\rightarrow T_{\varphi_t(X)}\mathcal{S}$.
The adjoint of $\mathbf{F}$ is defined as $\mathbf F^{\mathsf{T}}(X,t):T_{\varphi_t(X)}\mathcal{S}\rightarrow T_X\mathcal{B}$, $\mathbf g\left(\mathbf F \mathbf V,\boldsymbol{\mathrm{v}} \right)=\mathbf G\left(\mathbf{V},\mathbf F^{\mathsf{T}}\boldsymbol{\mathrm{v}}\right)$, $\forall\, \mathbf{V}\in T_X\mathcal{B},\, \boldsymbol{\mathrm{v}}\in T_{\varphi_t(X)}\mathcal{S}$.
The right Cauchy-Green deformation tensor is defined as
$\mathbf{C}(X,t)=\mathbf {F}^{\mathsf T}(X,t)\mathbf{F}(X,t):T_X\mathcal{B}\rightarrow T_X\mathcal{B}\,$. The Finger deformation tensor is defined as $\mathbf{b}(x,t)=\mathbf F(X,t)\mathbf{F}^{\mathsf{T}}(X,t):T_x\varphi\left(\mathcal{B}\right)\rightarrow T_x\varphi\left(\mathcal{B}\right)$, in components, $b^{ab}=F^a{}_AF^b{}_BG^{AB}$. Another measure of strain is the Lagrangian strain tensor given as $\mathbf E=\frac{1}{2}\left(\varphi^*_t\mathbf g-\mathbf G\right)$. The Jacobian of deformation $J$ relates the Riemannian volume element of the material manifold $dV(X,\mathbf G)$ to that of the spatial manifold $dv(\varphi_t(X),\mathbf g)$, written as
\begin{equation}
J=\sqrt{\frac{\mathrm{det}\,\mathbf g}{\mathrm{det}\,\mathbf G}}\,\mathrm{det}\,\mathbf F\,, \quad dv(x,\mathbf{g})=J\,dV(X,\mathbf{G})\,.
\end{equation}

\paragraph{Equilibrium Equations.}The localized balance of linear momentum in spatial and material forms are written as
\begin{equation}\label{linear-momentum}
 \operatorname{div}\boldsymbol{\sigma}+\rho\mathbf{b}=\rho\mathbf{a}, \quad
 \operatorname{Div}\mathbf{P}+\rho_0\mathbf{B}=\rho_0\mathbf{A},
\end{equation}
where $\boldsymbol{\sigma}$ and $\mathbf{P}$ are the Cauchy stress and the first Piola-Kirchhoff stress, respectively. Note that the material and spatial divergence operators in components are given as  
\begin{equation}
\begin{split}
\left(\mathrm{div}\boldsymbol \sigma\right)^a=&\sigma^{ab}{}_{|b}=\frac{\partial \sigma^{ab}}{\partial x^b}+\sigma^{ac}\gamma^b{}_{cb} +\sigma^{cb}\gamma^a{}_{cb}\,,\\
\left(\mathrm{Div} \mathbf P\right)^a=&P^{aA}{}_{|A}=\frac{\partial P^{aA}}{\partial X^A}+P^{aB}\Gamma^A{}_{AB} +P^{cA}F^b{}_A\gamma^a{}_{bc}\,,
\end{split}
\label{bal}
\end{equation}
where $\gamma^a{}_{bc}$ and $\Gamma^A{}_{BC}$ denote the Christoffel symbols of the connections $\nabla^{\mathbf{g}}$ and $\nabla^{\mathbf{G}}$, respectively. Note that in the local coordinate charts $\{x^a\}$ and $\{X^A\}$, one has ${\nabla^{\mathbf{g}}}_ {\partial_b}\partial_c=\gamma^a{}_{bc}\partial_a\,$ and ${\nabla^{\mathbf{G}}}_ {\partial_B}\partial_C=\Gamma^A{}_{BC}\partial_A\,$, respectively.

\paragraph{Constitutive Equations.} In this paper our calculations are restricted to incompressible transversely isotropic, orthotropic, and monoclinic solids. To establish a materially covariant strain energy density function, \emph{structural tensors} corresponding to the symmetry group of the material are used. For detailed discussions on structural tensors and the determination of the integrity basis and the corresponding invariants of a set of tensors, see \citep{spencer1971part,spencer1982formulation,liu1982representations,zheng1993tensors,lu2000covariant}.

\paragraph{Transverse Isotropy.} Let us assume a compressible transversely isotropic material such that the unit vector $\mathbf{N}(\mathrm X)$ identifies the material preferred direction at a point $\mathrm X$ in the reference configuration. The strain energy density per unit volume of the reference configuration is given as (see, e.g., \citep{doyle1956nonlinear,spencer1982formulation,lu2000covariant}) $W=W(\mathrm {X},\mathbf{G},\mathbf{C}^\flat, \mathbf{A})$, where $\mathbf{A}=\mathbf{N}\otimes\mathbf{N}$ is a structural tensor representing the transverse isotropy of the material symmetry group. The second Piola-Kirchhoff stress tensor is given by 
\begin{equation}
\mathbf {S}=2\frac{\partial W}{\partial \mathbf{C}^\flat}\,.
\label{secpiol}
\end{equation}
The energy function $W$ depends on the following five independent invariants defined as 
\begin{equation}
I_1=\mathrm{tr}\,\mathbf{C}\,,\quad I_2=\mathrm{det}\,\mathbf{C}\,\mathrm{tr}~\mathbf{C}^{-1}\,,\quad I_3=\mathrm{det}\,\mathbf{C}\,,\quad I_4=\mathbf{N}\cdot\mathbf{C}\cdot\mathbf{N}\,,\quad I_5=\mathbf{N}\cdot\mathbf{C}^2\cdot\mathbf{N}\,.
\label{invartr}
\end{equation}
In components they read
\begin{equation}
I_1=C^A{}_A\,,~ I_2=\mathrm{det}(C^A{}_B)(C^{-1})^D{}_D\,,~ I_3=\mathrm{det}(C^A{}_B)\,,~ I_4=N^AN^BC_{AB}\,,~  I_5=N^AN^BC_{BQ}C^Q{}_A\,.
\end{equation}
Using \eqref{secpiol}, one obtains\footnote{For the sake of brevity, we do not assume an explicit dependence of $W$ on $\mathrm{X}$, which in the case of inhomogeneous bodies is needed. We suppose instead that the material is piece-wise homogeneous and model an inhomogeneity using different energy functions in different regions of the body.}
\begin{equation} \label{piolacomt}
	\mathbf{S}=\sum_{n=1}^{5}2W_{I_n}\frac{\partial I_n}{\partial \mathbf{C}^\flat}\,,\qquad W_{I_n}
	:=\frac{\partial W}{\partial I_n}\,,\,\,~ n=1,\dots,5\,.
\end{equation}
Note that
\begin{equation}
\frac{\partial I_1}{\partial\mathbf{C}^\flat}=\mathbf{G}^{\sharp}\,,~~ \frac{\partial I_2}{\partial\mathbf{C}^\flat}=I_2\mathbf{C}^{-1}-I_3\mathbf{C}^{-2}\,,~~
\frac{\partial I_3}{\partial\mathbf{C}^\flat}=I_3\mathbf{C}^{-1}
\,,~~
\frac{\partial I_4}{\partial\mathbf{C}^\flat}=\mathbf{N}\otimes\mathbf{N}\,,\quad\,\frac{\partial I_5}{\partial\mathbf{C}^\flat}=\mathbf{N}\otimes\mathbf{C}\cdot\mathbf{N}+\mathbf{N}\cdot\mathbf{C}\otimes\mathbf{N}\,.
\label{invar}
\end{equation}
Thus, from \eqref{piolacomt} and \eqref{invar}, one obtains the following representation for the second Piola-Kirchhoff stress tensor
\begin{equation}
\mathbf{S}=2\left\{W_{I_1}\mathbf{G}^\sharp+W_{I_2}\left(I_2\mathbf{C}^{-1}-I_3\mathbf{C}^{-2}\right)+W_{I_3}I_3\mathbf{C}^{-1}+W_{I_4}\left(\mathbf{N}\otimes\mathbf{N}\right)+W_{I_5}\left(\mathbf{N}\otimes\mathbf{C}\cdot\mathbf{N}+\mathbf{N}\cdot\mathbf{C}\otimes\mathbf{N}\right)\right\}\,.
\label{piolas}
\end{equation} 
If the material is incompressible, then $I_3=1$, and thus, $W=W\left(\mathrm{X},I_1,I_2,I_4,I_5\right)$. Therefore, from \eqref{piolas}, $\mathbf{S}$ is expressed as
\begin{equation}
\mathbf{S}=2\left\{W_{I_1}\mathbf{G}^\sharp+W_{I_2}\left(I_2\mathbf{C}^{-1}-\mathbf{C}^{-2}\right)+W_{I_4}\left(\mathbf{N}\otimes\mathbf{N}\right)+W_{I_5}\left(\mathbf{N}\otimes\mathbf{C}\cdot\mathbf{N}+\mathbf{N}\cdot\mathbf{C}\otimes\mathbf{N}\right)\right\}-p\mathbf{C}^{-1}\,,
\end{equation} 
in which $p$ is the Lagrange multiplier associated with the incompressibility condition $J=1$. The Cauchy stress tensor $\sigma^{ab}=\frac{1}{J}F^a{}_A F^b{}_BS^{AB}$ is represented in component form as\footnote{Note that one can use the Cayley-Hamilton theorem and obtain
\begin{equation}
\frac{\partial I_{2}}{\partial\mathbf{C}^\flat}=I_{2}(\mathbf{C}^{-1})^\sharp-I_3(\mathbf{C}^{-2})^\sharp=I_1\mathbf{G}^\sharp-\mathbf{C}^\sharp.
\end{equation}}
\begin{equation}
\sigma^{ab}=2 F^a{}_A F^b{}_B \left[(W_{I_1}+I_1W_{I_2})G^{AB}-W_{I_2}C^{AB}+W_{I_4}N^A N^B+W_{I_5}\left(N^QN^AC^B{}_Q+N^PN^BC_P{}^A\right)\right]-pg^{ab}.
\label{cauchytr}
\end{equation}  

\paragraph{Orthotropy.} Next, we consider a compressible orthotropic material with three $\mathbf G$-orthonormal vectors $\mathbf{N}_1(\mathrm{X})$, $\mathbf{N}_2(\mathrm{X})$, and $\mathbf{N}_3(\mathrm{X})$ specifying the orthotropic axes in the reference configuration at a point $\mathrm{X}$. A choice of structural tensors is given by $\mathbf{A}_1=\mathbf{N}_1\otimes\mathbf{N}_1$, $\mathbf{A}_2=\mathbf{N}_2\otimes\mathbf{N}_2$, and $\mathbf{A}_3=\mathbf{N}_3\otimes\mathbf{N}_3$, where only two of which are independent as $\mathbf{A}_1+\mathbf{A}_2+\mathbf{A}_3=\mathbf{I}$. Hence, the energy function is given as \citep{doyle1956nonlinear,spencer1982formulation,lu2000covariant}
\begin{equation}\label{mon-ort}
W=W(\mathrm {X},\mathbf{G},\mathbf{C}^\flat, \mathbf{A}_1,\mathbf{A}_2)\,.
\end{equation}
The energy function $W$ is represented in terms of the following seven independent invariants
\begin{equation}
\begin{split}
I_1=\mathrm{tr}\,\mathbf{C}\,,\quad I_2=\mathrm{det}\,\mathbf{C}\,\mathrm{tr}~\mathbf{C}^{-1}\,,\quad I_3=\mathrm{det}\,\mathbf{C}\,,\quad I_4=\mathbf{N}_1\cdot\mathbf{C}\cdot\mathbf{N}_1\,,\\ I_5=\mathbf{N}_1\cdot\mathbf{C}^2\cdot\mathbf{N}_1\,,\quad I_6=\mathbf{N}_2\cdot\mathbf{C}\cdot\mathbf{N}_2\,,\quad I_7=\mathbf{N}_2\cdot\mathbf{C}^2\cdot\mathbf{N}_2\,.
\end{split}
\label{invaror}
\end{equation}
Using \eqref{secpiol}, one obtains
\begin{equation}
	\mathbf{S}=\sum_{n=1}^{7}2W_{I_n}\frac{\partial I_n}{\partial \mathbf{C}^\flat}\,,\qquad W_{I_n}
	:=\frac{\partial W}{\partial I_n}\,,\,\,~ n=1,\dots,7\,.
\label{piolas2}
\end{equation}
Substituting \eqref{invar} into \eqref{piolas2}, the second Piola-Kirchhoff stress tensor is given by
\begin{multline}
\mathbf{S}=2\Big\{W_{I_1}\mathbf{G}^\sharp+W_{I_2}\left(I_2\mathbf{C}^{-1}-I_3\mathbf{C}^{-2}\right)+W_{I_3}I_3\mathbf{C}^{-1}+W_{I_4}\left(\mathbf{N}_1\otimes\mathbf{N}_1\right)+W_{I_5}\left(\mathbf{N}_1\otimes\mathbf{C}\cdot\mathbf{N}_1+\mathbf{N}_1\cdot\mathbf{C}\otimes\mathbf{N}_1\right)\\+W_{I_6}\left(\mathbf{N}_2\otimes\mathbf{N}_2\right)+W_{I_7}\left(\mathbf{N}_2\otimes\mathbf{C}\cdot\mathbf{N}_2+\mathbf{N}_2\cdot\mathbf{C}\otimes\mathbf{N}_2\right)\Big\}\,.~~~~~
\label{piolacomor}
\end{multline}
In the case of incompressible solids $I_3=1$ and $W=W\left(\mathrm{X},I_1,I_2,I_4,I_5,I_6,I_7\right)$. Therefore, using \eqref{piolacomor}, one obtains the following representation for the second Piola-Kirchhoff stress tensor
\begin{multline}
\mathbf{S}=2\Big\{W_{I_1}\mathbf{G}^\sharp+W_{I_2}\left(I_2\mathbf{C}^{-1}-\mathbf{C}^{-2}\right)+W_{I_4}\left(\mathbf{N}_1\otimes\mathbf{N}_1\right)+W_{I_5}\left(\mathbf{N}_1\otimes\mathbf{C}\cdot\mathbf{N}_1+\mathbf{N}_1\cdot\mathbf{C}\otimes\mathbf{N}_1\right)\\+W_{I_6}\left(\mathbf{N}_2\otimes\mathbf{N}_2\right)+W_{I_7}\left(\mathbf{N}_2\otimes\mathbf{C}\cdot\mathbf{N}_2+\mathbf{N}_2\cdot\mathbf{C}\otimes\mathbf{N}_2\right)\Big\}-p\mathbf{C}^{-1}\,.~~~~~~~~~~
\end{multline}
In components, the Cauchy stress tensor is given as
\begin{multline}
\sigma^{ab}=2 F^a{}_A F^b{}_B \Big[(W_{I_1}+I_1W_{I_2})G^{AB}-W_{I_2}C^{AB}+W_{I_4}{N_1}^A {N_1}^B+W_{I_5}\left({N_1}^Q{N_1}^AC^B{}_Q+{N_1}^P{N_1}^BC_P{}^A\right)\\+W_{I_6}{N_2}^A {N_2}^B+W_{I_7}\left({N_2}^S{N_2}^AC^B{}_S+{N_2}^K{N_2}^BC_K{}^A\right)\Big]-pg^{ab}\,.
\label{cauchyor}
\end{multline}

\paragraph{Monoclinic Symmetry.} One of the preferred directions of a material with a monoclinic symmetry (say $\mathbf{N}_3(\mathrm X)$) is perpendicular to the plane of the other two (denoted by $\mathbf{N}_1(\mathrm X)$ and $\mathbf{N}_2(\mathrm X)$), which are not orthogonal. As an example one can consider an isotropic base material reinforced with two families of fibers such that the fibers are not at right angles, nor are they mechanically equivalent. In this case, the energy function is similar to that of orthotropic materials given by \eqref{mon-ort}, where $\mathbf{A}_1=\mathbf{N}_1\otimes\mathbf{N}_1$ and $\mathbf{A}_2=\mathbf{N}_2\otimes\mathbf{N}_2$. Nonetheless, an extra invariant $I_8=(\mathbf{N}_1\cdot\mathbf{N}_2)\mathbf{N}_1\cdot\mathbf{C}\cdot\mathbf{N}_2$ that models the coupling between the fibers (in $\mathbf{N}_1$ and $\mathbf{N}_2$ directions) 
is needed to express the energy function for monoclinic materials as $\mathbf{N}_1$ and $\mathbf{N}_2$ are not perpendicular (see \citep{MERODIO200,vergori2013anisotropic,demirkoparan}). Therefore   
\begin{equation}
	\mathbf{S}=\sum_{n=1}^{8}2W_{I_n}\frac{\partial I_n}{\partial \mathbf{C}^\flat}\,,\qquad W_{I_n}
	:=\frac{\partial W}{\partial I_n}\,,\,\,~ n=1,\dots,8\,.
\label{piolas2-mo}
\end{equation}
Hence\footnote{Note that $\frac{\partial I_8}{\partial\mathbf{C}^\flat}=\mathbf{N}_1\otimes\mathbf{N}_2+\mathbf{N}_2\otimes\mathbf{N}_1$.}
\begin{multline}
\mathbf{S}=2\Big\{W_{I_1}\mathbf{G}^\sharp+W_{I_2}\left(I_2\mathbf{C}^{-1}-I_3\mathbf{C}^{-2}\right)+W_{I_3}I_3\mathbf{C}^{-1}+W_{I_4}\left(\mathbf{N}_1\otimes\mathbf{N}_1\right)+W_{I_5}\left(\mathbf{N}_1\otimes\mathbf{C}\cdot\mathbf{N}_1+\mathbf{N}_1\cdot\mathbf{C}\otimes\mathbf{N}_1\right)\\+W_{I_6}\left(\mathbf{N}_2\otimes\mathbf{N}_2\right)+W_{I_7}\left(\mathbf{N}_2\otimes\mathbf{C}\cdot\mathbf{N}_2+\mathbf{N}_2\cdot\mathbf{C}\otimes\mathbf{N}_2\right)+\frac{W_{I_8}}{2}\left(\mathbf{N}_1\otimes\mathbf{N}_2+\mathbf{N}_2\otimes\mathbf{N}_1\right)\Big\}\,,~~~~~
\label{piola-mono}
\end{multline}
and for incompressible solids
\begin{multline}
\begin{split}
\mathbf{S}=&2\Big\{W_{I_1}\mathbf{G}^\sharp+W_{I_2}\left(I_2\mathbf{C}^{-1}-\mathbf{C}^{-2}\right)+W_{I_4}\left(\mathbf{N}_1\otimes\mathbf{N}_1\right)+W_{I_5}\left(\mathbf{N}_1\otimes\mathbf{C}\cdot\mathbf{N}_1+\mathbf{N}_1\cdot\mathbf{C}\otimes\mathbf{N}_1\right)\\&+W_{I_6}\left(\mathbf{N}_2\otimes\mathbf{N}_2\right)+W_{I_7}\left(\mathbf{N}_2\otimes\mathbf{C}\cdot\mathbf{N}_2+\mathbf{N}_2\cdot\mathbf{C}\otimes\mathbf{N}_2\right)+\frac{W_{I_8}}{2}\left(\mathbf{N}_1\otimes\mathbf{N}_2+\mathbf{N}_2\otimes\mathbf{N}_1\right)\Big\}-p\mathbf{C}^{-1}\,.
\end{split}
\end{multline}
The Cauchy stress is given in components as
\begin{multline}
\begin{split}
\sigma^{ab}&=2 F^a{}_A F^b{}_B \Big[(W_{I_1}+I_1W_{I_2})G^{AB}-W_{I_2}C^{AB}+W_{I_4}{N_1}^A {N_1}^B+W_{I_5}\left({N_1}^Q{N_1}^AC^B{}_Q+{N_1}^P{N_1}^BC_P{}^A\right)\\&+W_{I_6}{N_2}^A {N_2}^B+W_{I_7}\left({N_2}^S{N_2}^AC^B{}_S+{N_2}^K{N_2}^BC_K{}^A\right)+\frac{W_{I_8}}{2}\left({N_1}^A {N_2}^B+{N_2}^A {N_1}^B\right)\Big]-pg^{ab}\,.
\end{split}
\label{cauchymon}
\end{multline}

\paragraph{Cartan's Moving Frame.}At a point $X$ of a manifold $\mathcal{B}$ consider an orthonormal frame field $\{\mathbf{e}_{\alpha}\}_{\alpha=1}^N$ forming a basis for $T_X\mathcal{B}$. This frame field is not necessarily a coordinate basis for the tangent space. However, given a coordinate basis $\{\frac{\partial}{\partial X^A}\}$, one can obtain an arbitrary frame field $\{\mathbf{e}_{\alpha}\}$ using an $SO(N,\mathbb{R})$-rotation of the coordinate basis such that $\mathbf{e}_{\alpha}=\mathsf{F}^A{}_{\alpha}\frac{\partial}{\partial X^A}$. For a coordinate frame $\left[\frac{\partial }{\partial X^A},\frac{\partial }{\partial X^B}\right]=0$,\footnote{Note that for any pair of vector fields $\mathbf{U}$ and $\mathbf{V}$ on $\mathcal{B}$, one can define a new vector field \textemdash the commutator \textemdash given by $[\mathbf{U},\mathbf{V}]_Xf:=\mathbf{U}_X(\mathbf{V}f)-\mathbf{V}_X(\mathbf{V}f)$, for any smooth function at $X$ on $\mathcal{B}$.} 
whereas for the non-coordinate frame, $[\mathbf{e}_{\alpha},\mathbf{e}_{\beta}]=-c^{\gamma}{}_{\alpha\beta}\mathbf{e}_\gamma$, where $c^{\gamma}{}_{\alpha\beta}$ are the componenets of the object of anhonolomy. 
One can show that $c^{\gamma}{}_{\alpha\beta}=\mathsf{F}^A{}_{\alpha}\mathsf{F}^B{}_{\beta}\left(\partial_A\mathsf{F}^{\gamma}{}_B-\partial_B\mathsf{F}^{\gamma}{}_A\right)$, where $\mathsf{F}^{\gamma}{}_A$ is the inverse of $\mathsf{F}^A{}_{\gamma}$. 
Connection $1$-forms are defined by $\nabla\mathbf{e}_{\alpha}=\mathbf{e}_{\gamma}\otimes\omega^{\gamma}{}_{\alpha}$, and in components, $\nabla_{\mathbf{e}_{\beta}}\mathbf{e}_{\alpha}=\left<\omega^{\gamma}{}_{\alpha},\mathbf{e}_{\beta}\right>\mathbf{e}_{\gamma}=\omega^{\gamma}{}_{\beta\alpha}\mathbf{e}_{\gamma}$. In terms of the co-frame field $\{\vartheta^\alpha\}_{\alpha=1}^N$ corresponding to $\{\mathbf{e}_\alpha\}$, one has $\omega^{\gamma}{}_\alpha=\omega^{\gamma}{}_{\beta\alpha}\vartheta^{\beta}$. Similarly, one obtains $\nabla\vartheta^{\alpha}=-\omega^{\alpha}{}_{\gamma}\vartheta^{\gamma}$ and $\nabla_{\mathbf{e}_{\beta}}\vartheta^{\alpha}=-\omega^{\alpha}{}_{\beta\gamma}\vartheta^{\gamma}$. The metric tensor is represented as $\mathbf{G}=\delta_{\alpha\beta}\vartheta^{\alpha}\otimes\vartheta^{\beta}$. Metric compatibility of $\nabla$ gives the following constraints on the connection 1-forms $\delta_{\alpha\gamma}\omega^{\gamma}{}_{\beta}+\delta_{\beta\gamma}\omega^{\gamma}{}_{\alpha}=0$.
In a non-coordinate basis, the torsion and curvature have the following components $T^{\alpha}{}_{\beta\gamma} =\omega^{\alpha}{}_{\beta\gamma}-\omega^{\alpha}{}{}_{\gamma\beta}+c^{\alpha}{}_{\beta\gamma}$ and $\mathcal{R}^{\alpha}{}_{\beta\lambda\mu} =\partial_{\beta}\omega^{\alpha}{}_{\lambda\mu}-\partial_{\lambda}\omega^{\alpha}{}_{\beta\mu}+\omega^{\alpha}{}_{\beta\xi}\omega^{\xi}{}_{\lambda\mu}-\omega^{\alpha}{}_{\lambda\xi}\omega^{\xi}{}_{\beta\mu}+\omega^{\alpha}{}_{\xi\mu}c^{\xi}{}_{\beta\lambda}$, respectively.
Torsion and curvature $2$-forms are, respectively, given by $\mathcal{T}^{\alpha}=d\vartheta^{\alpha}+\omega^{\alpha}{}_{\beta}\wedge\vartheta^{\beta}$ and $\mathcal{R}^{\alpha}{}_{\beta} =d\omega^{\alpha}{}_{\beta}+\omega^{\alpha}{}_{\gamma}\wedge\omega^{\gamma}{}_{\beta}$.
These are called Cartan's first and second structural equations. The density of Burgers' vector $\mathbf{b}$ at a point $X$ of $\mathcal{B}$ is related to torsion 2-from as follows 
\begin{equation}\label{burger}
	b^{\alpha}(X;C_s)=\int_{\Omega_s}\mathsf{P}^{\alpha}{}_{\beta}\mathcal{T}^{\beta}\,,
\end{equation}
where $\Omega_s\in \mathcal{B}$ is a smooth surface with a boundary given by the curve $C_s$, and $\mathsf{P}(C_s)_{\tau}^t:T_{C_s(\tau)}\mathcal{B}\to T_{C_s(t)}\mathcal{B}$ parallel transports vectors tangent to the manifold at $C_s(\tau)$ to $C_s(t)$ (see \citep{katanaev2005,ozakin2014affine} for more details).

\section{Examples of Anisotropic Bodies with Distributed Defects}

In this section, we consider several examples of distributed defects in cylindrical bars made of orthotropic and monoclinic solids as well as distributed defects in spherical balls made of transversely isotropic solids. Particularly, we consider cylindrically-symmetric distributions of parallel screw dislocations and disclinations in an orthotropic medium, a spherically-symmetric distribution of point defects in a transversely isotropic spherical ball, and a cylindrically-symmetric distribution of screw dislocations in a monoclinic medium. We also discuss the effects of the constitutive parameters on the induced stress fields for different types of defects.  


\subsection{A Cylindrically-Symmetric Distribution of Parallel Screw Disclocations in an Orthotropic Medium}
\label{scr-or}

Let us consider a cylindrically-symmetric distribution of screw dislocations parallel to the $Z$-axis with a radially-symmetric Burgers' vector density $b(R)$ (in a cylindrical coordinate system $(R,\Theta,Z)$) in an infinite orthotropic medium. We assume that in the reference configuration the dislocated body is orthotropic. The material preferred directions at a material point $\mathrm X$ are denoted by $\mathbf{N}_1(\mathrm X)$, $\mathbf{N}_2(\mathrm X)$, and $\mathbf{N}_3(\mathrm X)$ in the reference configuration. In the current configuration, the preferred directions are given by $\mathbf{n}_1(\mathrm x)$, $\mathbf{n}_2(\mathrm x)$, and $\mathbf{n}_3(\mathrm x)$ at the ambient point $\mathrm x$ corresponding to the material point $\mathrm X$. We assume that $\mathbf{N}_1$ and $\mathbf{N}_2$ are in the radial and axial directions, respectively. Note that $\mathbf{N}_3$, which is perpendicular to $\mathbf{N}_1$ and $\mathbf{N}_2$, explicitly depends on the distribution of screw dislocations as will be seen in the following. This is because the geometry of the material manifold has an explicit nontrivial dependence on the dislocations distribution (see \eqref{_G}). In the current configuration, the body will have monoclinic anisotropy as $\mathbf{n}_1$ will be perpendicular to the plane of $\mathbf{n}_2$ and $\mathbf{n}_3$, which will not be orthogonal in the ambient space. 
It turns out that the material manifold for a nonlinear solid with distributed dislocations is a Weitzenb\"{o}ck manifold, i.e., a manifold with torsion having a flat connection and vanishing non-metricity (see \citep{YavariGoriely2012a,ozakin2014affine} for more details). Therefore, the material metric for the dislocated body is written as
\begin{equation}
\label{_G}
\mathbf{G}=
\left(
\begin{array}{ccc}
1&0&0\\
0&R^2+f(R)^2&f(R)\\
0&f(R)&1
\end{array}
\right)\,,
\end{equation}
where $f(R)$ is related to the Burgers' vector density $b(R)$ such that $f'(R)=\frac{R}{2\pi}b(R)$. 
Let us endow the ambient space with the Euclidean metric $\mathbf{g}=\operatorname{diag}\{1,r^2,1\}$. We then assume an embedding of the material manifold into the ambient space of the form $\left(r,\theta,z\right)=\left(r\left(R\right),\Theta,\alpha Z\right)$, where $\alpha$ is a positive constant denoting the longitudinal stretch. Hence, $\mathbf{F}=\operatorname{diag}\{r'(R),1,\alpha\}$. Assuming incompressibility, i.e., $J=\sqrt{\frac{\mathrm{det} \mathbf g}{\mathrm{det} \mathbf G}} \mathrm{det} \mathbf F=1$, one obtains $\frac{r(R)}{R}r'(R)\alpha=1$. Eliminating the rigid body translation by setting $r(0)=0$, one obtains $r(R)=\frac{1}{\sqrt{\alpha}}R$. Therefore, the right Cauchy-Green deformation tensor is written as\footnote{The symbolic computations in this paper were performed using Mathematica \citep{mathematicap}.}
\begin{equation}\label{disco-c}
\mathbf{C}=
\left(
\begin{array}{ccc}
\frac{1}{\alpha}&0&0\\\\
0&\frac{1}{\alpha}&-\frac{\alpha^2f(R)}{R^2}\\\\
0&-\frac{f(R)}{\alpha}&\frac{\alpha^2}{R^2}(R^2+f(R)^2)
\end{array}
\right)\,.
\end{equation}
Note that $\mathbf{N}_1=\mathbf{E}_R$, $\mathbf{N}_2=\mathbf{E}_Z$, and $\mathbf{N}_3=\frac{1}{R}\mathbf{E}_\Theta-\frac{f(R)}{R}\mathbf{E}_Z$. Note also that $\mathbf{N}_3$ is obtained using the orthonormality of the material preferred directions, and   $\mathbf{E}_R={\partial}/{\partial R}$, $\mathbf{E}_Z={\partial}/{\partial Z}$, and $\mathbf{E}_\Theta={\partial}/{\partial \Theta}$ form a basis for $T_X\mathcal{B}$. Using \eqref{invaror}, the invariants of the strain energy function are simplified and are written as
\begin{equation}\label{inv-or-dis}
\begin{aligned}
I_1=&\operatorname{tr}\mathbf{C}=\frac{2}{\alpha}+\frac{\alpha^2}{R^2}(R^2+f(R)^2)\,, \quad
I_2=
\frac{1}{2}\left[\operatorname{tr}(\mathbf{C}^2)-(\operatorname{tr}\mathbf{C})^2\right]=\frac{1}{\alpha^2}+2\alpha+\alpha\frac{f(R)^2}{R^2}\,,\\	
I_4=&\frac{1}{\alpha}\,,\qquad
I_5=\frac{1}{\alpha^2}\,,\qquad
I_6=\alpha^2\,, \quad
I_7= \frac{\alpha^4}{R^2}(R^2+f(R)^2) \,.
\end{aligned}
\end{equation}
The non-zero components of the Cauchy stress tensor following \eqref{cauchyor} read 
\begin{align}
\sigma^{rr}=&\frac{2}{\alpha ^2} \left[W_{I_2} \Big(\alpha
^3+\frac{\alpha ^3 f(R)^2}{R^2}+1\Big)+\alpha 
W_{I_4}+2 W_{I_5}\right]+\frac{2 W_{I_1}}{\alpha }-p(R)\,,\label{sirr2}  \\
\sigma^{\theta\theta}=&\frac{2 \alpha  W_{I_1}+2 \left(\alpha^3+1\right) W_{I_2}-\alpha^2p(R)}{\alpha  R^2}\,,\label{siphi}\\
\sigma^{zz}=&\frac{2 \alpha}{R^2}  \Big[ \left(f(R)^2+R^2\right) (\alpha W_{I_1}+W_{I_2}+2 \alpha^3W_{I_7})+R^2 (W_{I_2}+\alpha   W_{I_6})\Big]-p(R)\,,\\
\sigma^{\theta z}=&-\frac{2 f(R)}{R^2} \left(\alpha  W_{I_1}+W_{I_2}+\alpha ^3 W_{I_7}\right)\,.
\label{tetateta}
\end{align}
We assume that the stress vanishes when the body is dislocation-free and the longitudinal stretch $\alpha=1$ (see also \citep{merodio2003instabilities,vergori2013anisotropic,Golgoon2017}). Thus
\begin{equation}
\left(W_{I_4}+2W_{I_5}\right)|_{I_1=I_2=3, I_4=I_5=I_6=I_7=1}=0\,,\quad\mathrm{and}\quad \left(W_{I_6}+2W_{I_7}\right)|_{I_1=I_2=3, I_4=I_5=I_6=I_7=1}=0\,.
\label{compor1}
\end{equation}
In the absence of body and inertial forces, the only non-trivial equilibrium equation is $\sigma^{rb}{}_{|b}=0$, implying\footnote{Note that $p=p(R)$ is implied from the other equilibrium equations.} that (cf. \eqref{bal}) $\sigma^{rr}{}_{,r}+\frac{\sigma^{rr}}{r}-r\sigma^{\theta\theta}=0$. Therefore, $p'(R)=h(R)$, where
\begin{multline}\label{hr1}
\begin{aligned}
h(R)=&\frac{2 }{\alpha  R^5}  \Big[2 R^3 f(R) f'(R) \Big(\alpha
	^2 W_{I_2}+\alpha ^2 W_{I_1I_1}+\alpha
	 (2+\alpha^3)W_{I_1I_2} +\alpha ^2 W_{I_1I_4}+2 \alpha
	W_{I_1I_5}+\alpha ^4 W_{I_1I_7}\\&+(\alpha
	^3+1) W_{I_2I_2}+\alpha 
	W_{I_2I_4}+2 W_{I_2I_5}+\alpha ^3(\alpha ^3+1)
	W_{I_2I_7}+\alpha
	^4 W_{I_4I_7}+2 \alpha ^3
	W_{I_5I_7}\Big)\\&+2 \alpha ^3 R f(R)^3 f'(R)
	\left(\alpha 
	W_{I_1I_2}+W_{I_2I_2}+\alpha ^3
	W_{I_2I_7}\right)-R^2 f(R)^2 \Big\{2 \alpha  W_{I_1I_2}(2+\alpha^3)\\&+2 \alpha ^2
	W_{I_1I_4}+4 \alpha  W_{I_1I_5}+2 \alpha
	^4 W_{I_1I_7}+\alpha ^2 W_{I_2}+2(\alpha ^3+1)
	 W_{I_2I_2}+2
	\alpha  W_{I_2I_4}+4 W_{I_2I_5}\\&+2 \alpha
	^3 (\alpha ^3+1)W_{I_2I_7}+2
	\alpha ^4 W_{I_4I_7}+4 \alpha ^3
	W_{I_5I_7}+2 \alpha ^2
	W_{I_1I_1}\Big\}\\&-2 \alpha ^3 f(R)^4
	\left(\alpha 
	W_{I_1I_2}+W_{I_2I_2}+\alpha ^3
	W_{I_2I_7}\right)+R^4 W_{I_4}\Big]+\frac{4
	 W_{I_5}}{\alpha ^2 R}\,.
\end{aligned}
\end{multline}
If one assumes that the medium is a cylinderical bar with a finite radius $R_o$ and the surface $R=R_o$ is traction-free, one obtains
\begin{equation}\label{press}
p(R)=\int_{R_o}^{R}h(\zeta)d\zeta+\frac{2}{\alpha ^2} \left[ \Big(\alpha
^3+\frac{\alpha ^3 f(R_o)^2}{R_o^2}+1\Big)W_{I_2}|_{R=R_o}+\alpha 
W_{I_4}|_{R=R_o}+2 W_{I_5}|_{R=R_o}\right]+\frac{2}{\alpha } W_{I_1}|_{R=R_o}\,.
\end{equation}
Let us employ the so called \emph{standard reinforcing model} for compressible materials, which is defined as \citep{triantafy,merodio2003instabilities,merodio2005tensile}
\begin{equation}
W=W\left(I_1,I_2,I_4,I_5,I_6,I_7\right)=W_{\mathrm{iso}}\left(I_1,I_2\right)+W_{\mathrm{fib}}^R\left(I_4,I_5\right)+W_{\mathrm{fib}}^Z\left(I_6,I_7\right)\,,
\label{reimo}
\end{equation}
where $W_{\mathrm{iso}}$ denotes the strain energy function for the isotropic base material, whereas $W_{\mathrm{fib}}^R$ and $W_{\mathrm{fib}}^Z$ represent the anisotropic effects due to the fiber reinforcement in the radial and longitudinal directions, respectively. Consider as an example a cylindrical body made of a Mooney-Rivlin solid reinforced with fibers in the radial and longitudinal directions such that 
\begin{equation}
\begin{aligned}
W(I_1,I_2,I_4,I_5,I_6,I_7) =&\frac{\mu_1}{2}\left(I_1-3\right)+\frac{\mu_2}{2}\left(I_2-3\right)+\frac{\gamma_1}{2}\left(I_4-1\right)^2\\
&+\frac{\gamma_2}{2}\left(I_5-1\right)^2+\frac{\xi_1}{2}\left(I_6-1\right)^2+\frac{\xi_2}{2}\left(I_7-1\right)^2\,.
\label{sten0}
\end{aligned}
\end{equation} 
Using \eqref{hr1}, we have
\begin{equation}
	h(R)=\alpha  \mu_2 \frac{f(R)}{R^3}\left[ 2  R  f'(R)- f(R)\right]+\frac{2}{R}\frac{1}{\alpha^2}
	\left(\frac{1}{\alpha}-1\right) \left [\alpha  \gamma_1 +2 \gamma_2
	\left(1+\frac{1}{\alpha}\right)\right]\,.
\end{equation}
Thus, from \eqref{press}
\begin{equation}\label{pr-dislo}
\begin{split}
	p(R)=&\alpha \mu_2\int_{R_o}^{R}\frac{f(\zeta)}{\zeta^3}\left[ 2  \zeta  f'(\zeta)- f(\zeta)\right]d\zeta+\frac{2}{\alpha^2}
	\left(\frac{1}{\alpha}-1\right) \left[\alpha  \gamma_1 
	+2 \gamma_2\left(1+\frac{1}{\alpha}\right)\right]
	\left[1+\ln\frac{R}{R_o}\right]\\&+ \frac{\mu_1}{\alpha }
	+\frac{\mu_2}{\alpha ^2}  \Big(\alpha^3+\frac{\alpha ^3 f(R_o)^2}{R_o^2}+1\Big) \,.
\end{split}
\end{equation}
The physical components of the Cauchy stress read\footnote{The physical components of the Cauchy stress tensor, i.e., ${\hat{\sigma}}^{ab}=\sigma^{ab}\sqrt{g_{aa}g_{bb}}$ (no summation) \citep{truesdell1953physical} are given as $\hat{\sigma}^{rr}=\sigma^{rr}$, $\hat{\sigma}^{\theta\theta}=r^2(R)\sigma^{\theta\theta}$, $\hat{\sigma}^{zz}=\sigma^{zz}$, and $\hat{\sigma}^{\theta z}=r(R)\sigma^{\theta z}$.
}
\begin{align}
\begin{split}
\hat{\sigma}^{rr}=&\alpha\mu_2 \Big(\frac{ f(R)^2}{R^2}-\frac{ f(R_o)^2}{R_o^2}\Big)+\alpha \mu_2\int_{R}^{R_o}\frac{f(\zeta)}{\zeta^3}\left[ 2  \zeta  f'(\zeta)- f(\zeta)\right]d\zeta\\&-\frac{2}{\alpha^2}
(\frac{1}{\alpha}-1) \left\{\alpha  \gamma_1 +2 \gamma_2
(1+\frac{1}{\alpha})\right\}\ln\frac{R}{R_o}\,,\label{psirr2} 
\end{split}
 \\
\begin{split}
\hat{\sigma}^{\theta\theta}=&\alpha \mu_2\int_{R}^{R_o}\frac{f(\zeta)}{\zeta^3}\left[ 2  \zeta  f'(\zeta)- f(\zeta)\right]d\zeta -\alpha\mu_2\frac{f(R_o)^2}{R_o^2}  \\ &-\frac{2}{\alpha^2}
(\frac{1}{\alpha}-1) \left\{\alpha  \gamma_1 +2 \gamma_2
\left(1+\frac{1}{\alpha}\right)\right\}\left[1+\ln\frac{R}{R_o}\right]\,,\label{psiphi}
\end{split}
\\
\begin{split}
\hat{\sigma}^{zz}=&2\alpha^2(\alpha^2-1)\xi_1+\frac{4\alpha^4\xi_2}{R^2}\left(f(R)^2+R^2\right) \Big[  \frac{\alpha^4}{R^2}\left(f(R)^2+R^2\right)-1\Big]\\&+\alpha\mu_2 \Big(\frac{ f(R)^2}{R^2}-\frac{ f(R_o)^2}{R_o^2}\Big)-\frac{2}{\alpha^2}
\left(\frac{1}{\alpha}-1\right) \left\{\alpha  \gamma_1 +2 \gamma_2
\left(1+\frac{1}{\alpha}\right)\right\}\left[1+\ln\frac{R}{R_o}\right]\\&+\alpha \mu_2\int_{R}^{R_o}\frac{f(\zeta)}{\zeta^3}\left[ 2  \zeta  f'(\zeta)- f(\zeta)\right]d\zeta+(\alpha\mu_1+\mu_2)(\alpha-\frac{1}{\alpha^2})+\alpha^2\mu_1\frac{f(R)^2}{R^2}\,,
\end{split}
\\
\hat{\sigma}^{\theta z}=&-\frac{ f(R)}{\alpha^{\frac{1}{2}}R} \left(\alpha \mu_1 +\mu_2+2\alpha ^3 \xi_2\left[\frac{\alpha^4}{R^2}\left(R^2+f(R)^2\right)-1\right]\right)\,.
\label{ptetateta}
\end{align}

\begin{remark}\label{sin-sc}
From \eqref{pr-dislo}, for an arbitrary cylindrically-symmetric distribution of parallel screw dislocations, the pressure $p(R)$, and hence, $\hat{\sigma}^{rr}$, $\hat{\sigma}^{\theta\theta}$, and $\hat{\sigma}^{zz}$ exhibit a logarithmic singularity on the dislocation axis $(R=0)$ unless $\alpha=1$. Note that this singularity is inherent to the anisotropic effects due to the reinforcement in the radial direction. In particular, the singularity does not occur when $\gamma_1=\gamma_2=0$, e.g., when the material is isotropic. 
\end{remark}

\begin{remark}
Note that in the case of fiber-reinforced neo-Hookean materials ($\mu_2=0$) and a given arbitrary cylindrically-symmetric distribution of screw dislocations supported on a cylinder of radius $R_i$, the stress field for $R>R_i$ is independent of $b(R)$ and is identical to that of a single screw dislocation with Burgers vector $b_0=\int_{0}^{R_i}\eta b(\eta)d\eta$. \citet{acharya2001model} and \citet{YavariGoriely2012a} observed that this result holds for isotropic neo-Hookean solids.
\end{remark}

\begin{figure}
	\centering
	\begin{center}
		\makebox[0pt]{\includegraphics[trim = 14.3cm 0.5cm 14.3cm 0.7cm, clip, scale=0.35]{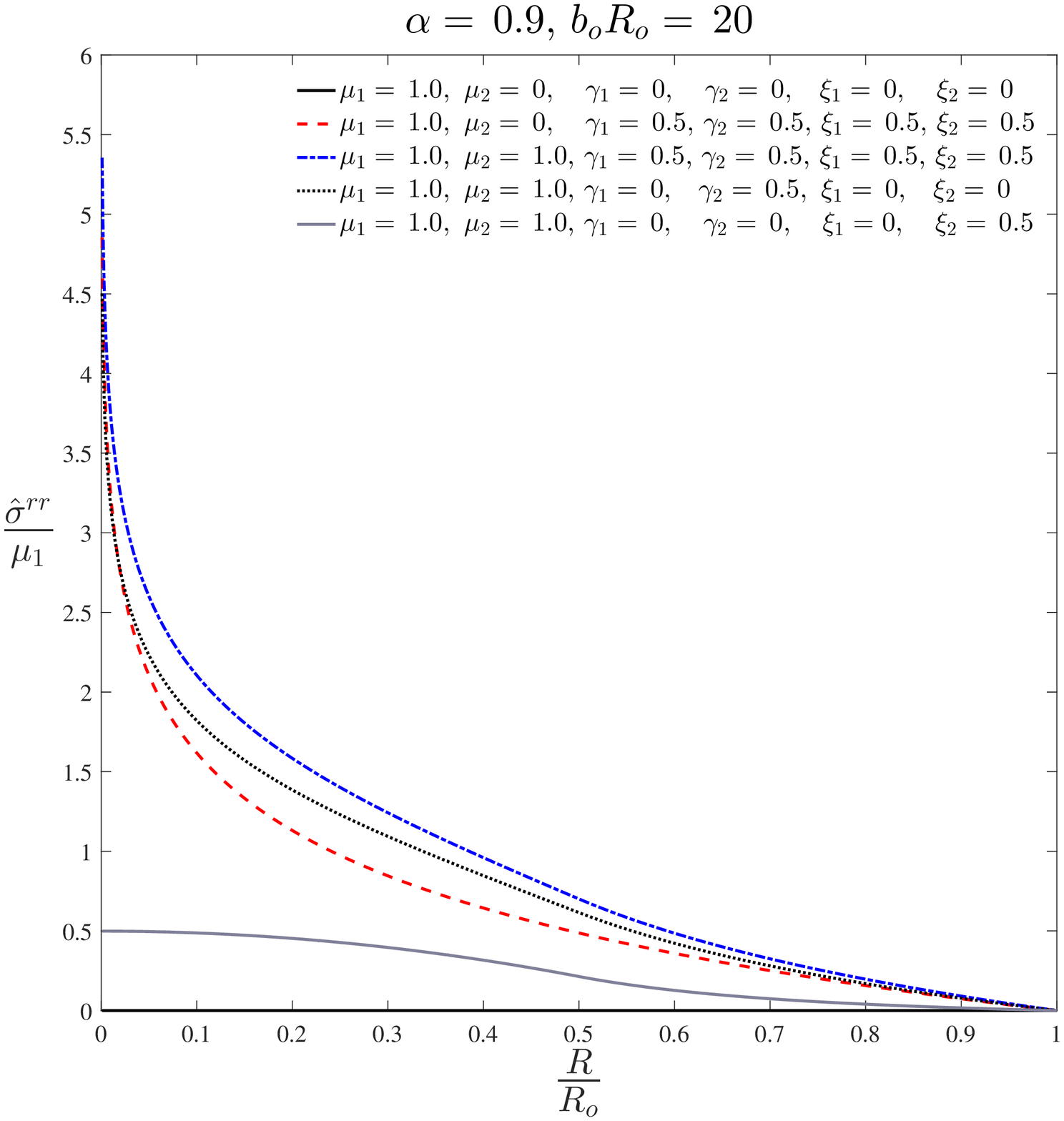}
			\includegraphics[trim = 14.3cm 0.5cm 14.3cm 0.7cm, clip, scale=0.35]{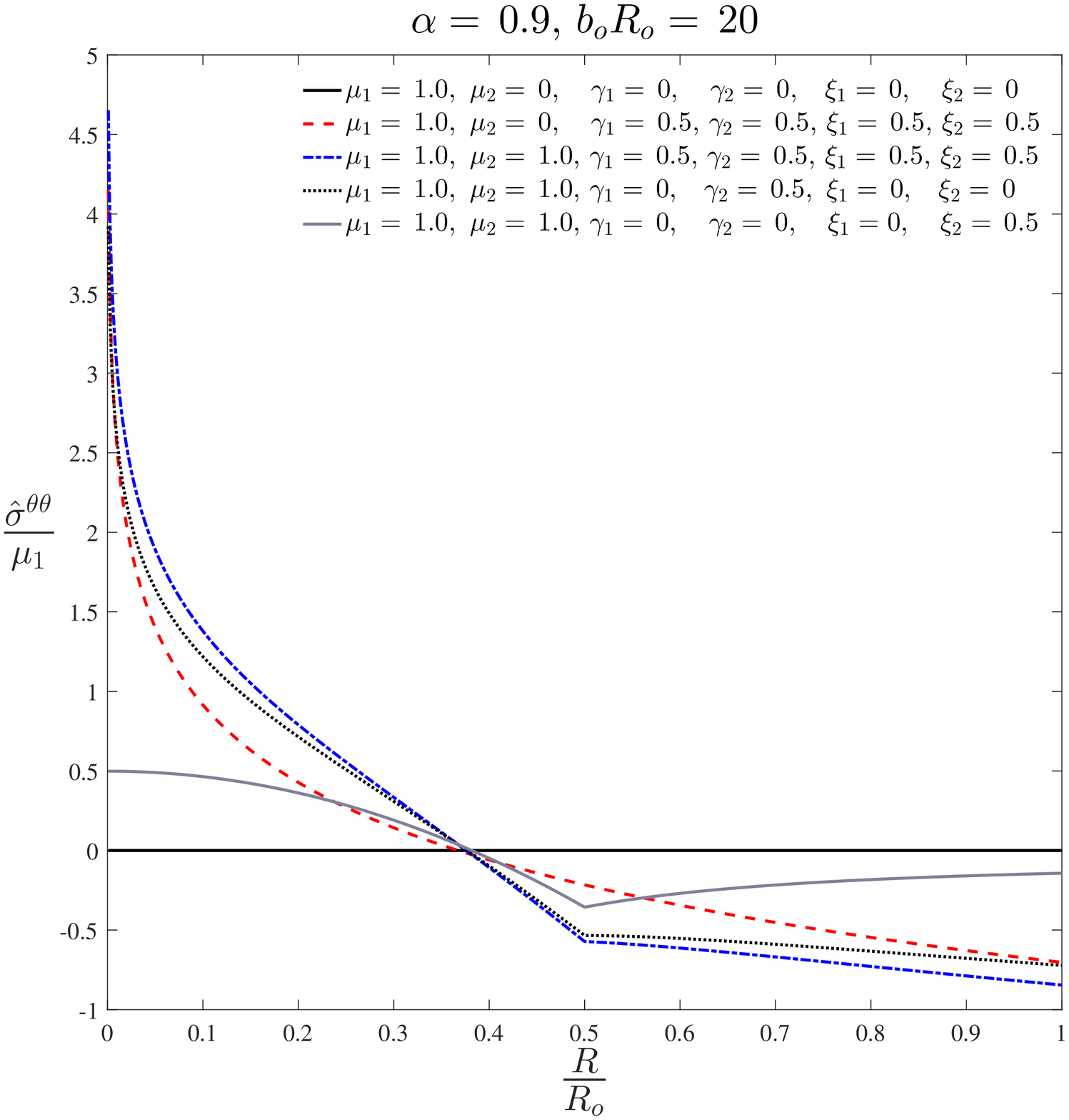}}
	\end{center}
	\begin{center}
		\makebox[0pt]{\includegraphics[trim = 14.3cm 0.5cm 14.3cm 0.7cm, clip, scale=0.35]{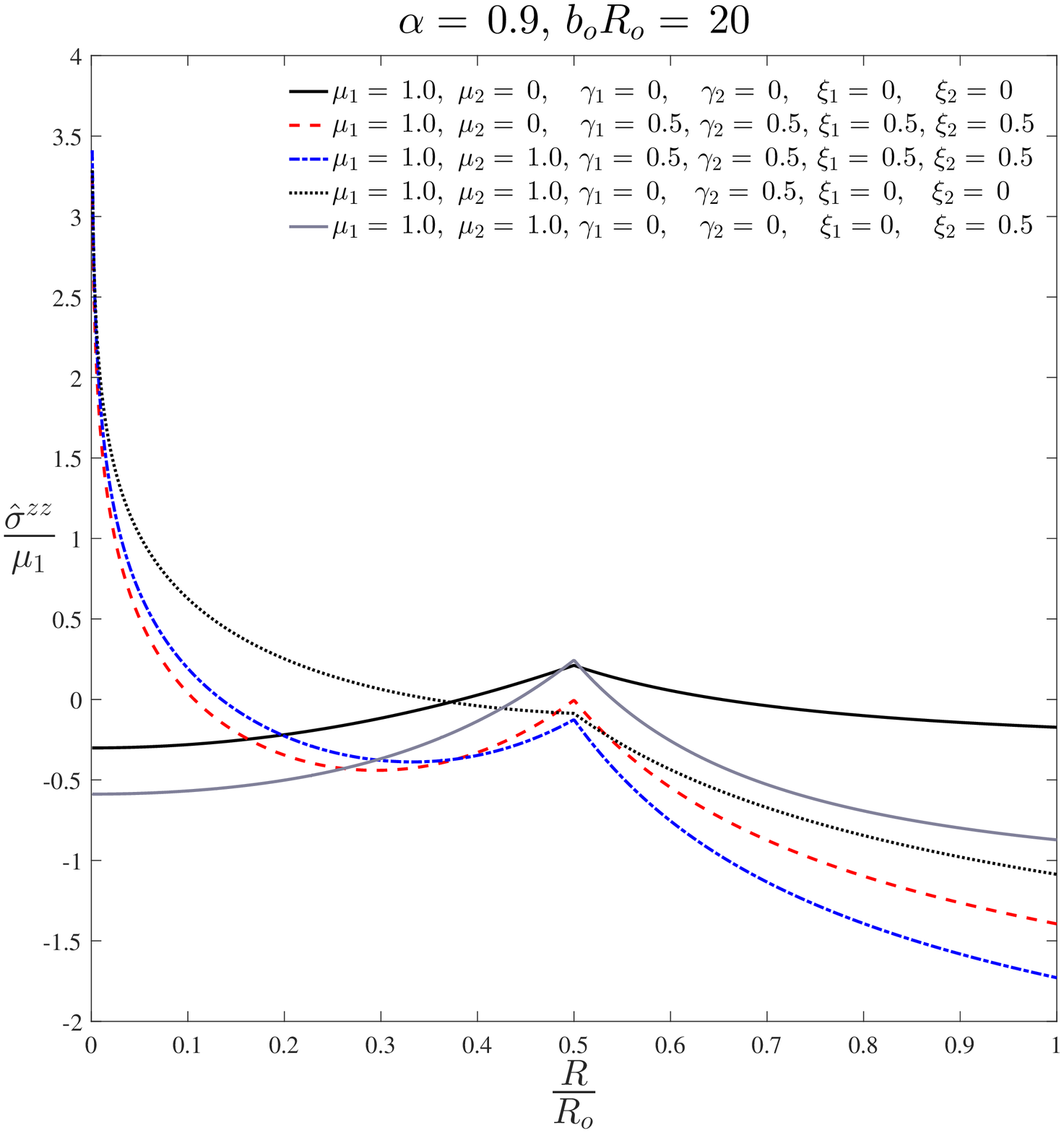}
			\includegraphics[trim = 14.3cm 0.5cm 14.3cm 0.7cm, clip, scale=0.35]{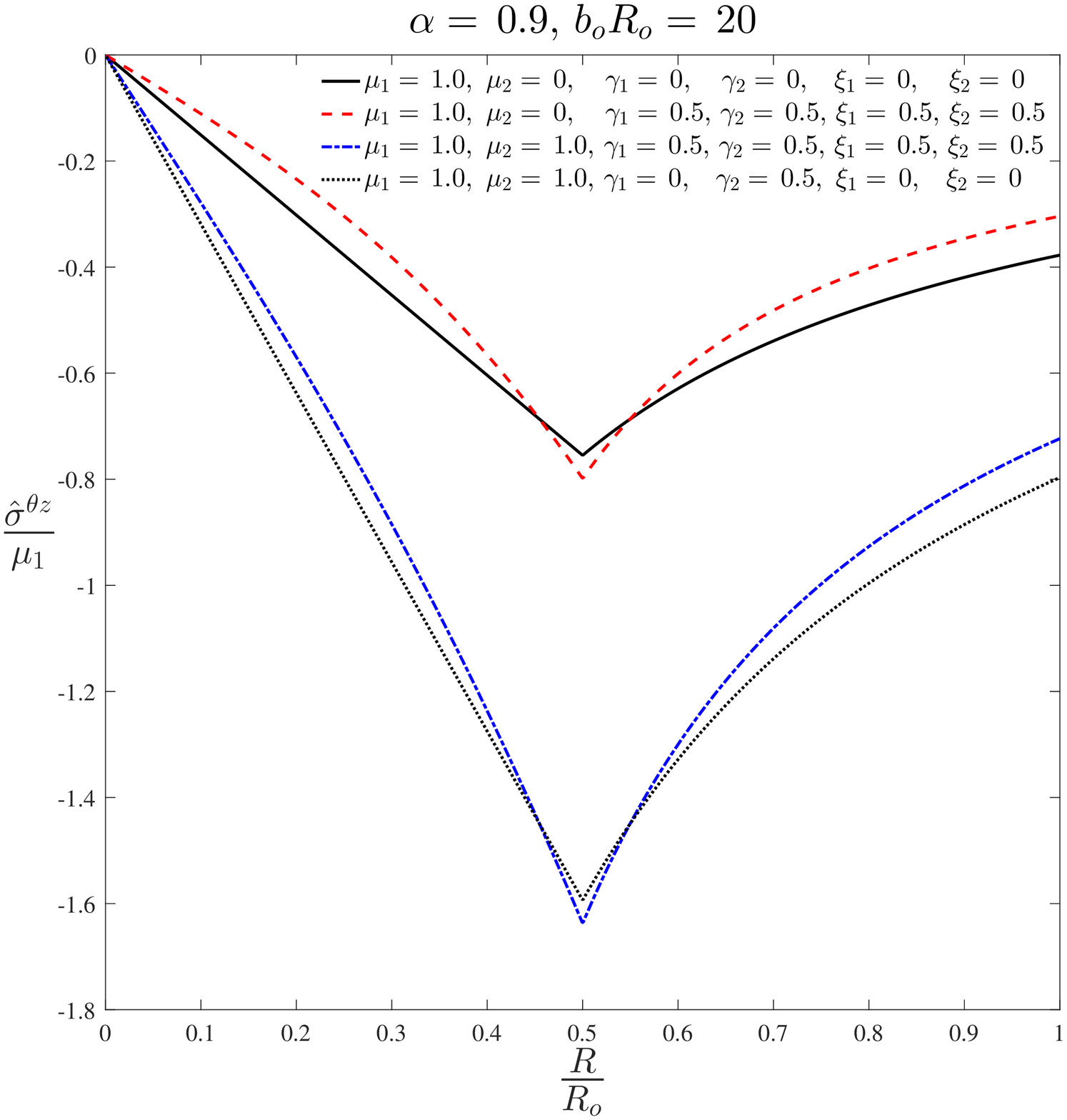}}
	\end{center}
	\vspace*{-0.2in}
	\caption{Stress distribution in a medium with the constitutive equation \eqref{sten0} and the dislocation distribution $\eqref{uni-bur}$ such that $R_i/R_o=0.5$, $b_0R_o=20$, and $\alpha=0.9$ for different values of the constitutive parameters.} 
	\label{sb}
\end{figure}

As an example, let us assume the following Burgers' vector density distribution:
\begin{equation}\label{uni-bur}
b(R)=
\begin{cases}
b_0\quad 0<R\leq R_i\,,\\
0\quad R_i<R\leq R_o\,,
\end{cases}
\end{equation}
where $R_i\leq R_o$. Thus
\begin{equation}
f(R)=\frac{1}{2\pi}\int_{0}^{R}\eta b(\eta)d\eta=\frac{b_0}{4\pi}
\begin{cases}
R^2\quad ~0<R\leq R_i\,,\\
R_i^2\quad R_i<R\leq R_o\,.
\end{cases}
\end{equation}
Fig.~\ref{sb} depicts the variation of the different components of the Cauchy stress for the Burgers' vector density distribution \eqref{uni-bur} such that $R_i/R_o=0.5$ and $b_0R_o=20$. Notice that the $\hat{\sigma}^{rr}$ and $\hat{\sigma}^{\theta\theta}$ vanish for a neo-Hookean solid.

\begin{remark}
As noted by \citet{zubov1997}, the energy per unit length (along the dislocation line) of a single screw dislocation in a Mooney-Rivlin solid is unbounded.\footnote{Note, however, that the energy of distributed screw dislocations is not necessarily unbounded (see also \citep{Sadik20160659}). In particular, a Mooney-Rivlin reinforced material with the energy function \eqref{sten0} and the Burgers' vector distribution \eqref{uni-bur} has a finite energy per unit length.} This is also the case for a fiber-reinforced Mooney-Rivlin material due to the standard reinforcing model considered here (cf. \eqref{reimo}). Let us consider incompressible isotropic base materials, for which the energy per unit length of a single screw dislocation remains bounded, i.e., $2\pi\int_{0}^{R_o}W_{\mathrm{iso}}(I_1(\xi),I_2(\xi))\xi d\xi<\infty$, for finite $R_o$ (examples include Varga \citep{zubov1997}, incompressible power-law \citep{knowles1977finite,rosakis1988screw}, generalized incompressible neo-Hookean \citep{YavariGoriely2012a} materials, and Hencky material \citep{Yavari2016}). Exploiting the standard reinforcing model, the energy function for the fiber-reinforced material with the isotropic base with the energy function $W_{\mathrm{iso}}(I_1,I_2)$ is assumed to be given as
\begin{equation}\label{rem-ren}
	W=W_{\mathrm{iso}}(I_1,I_2)+\frac{\gamma_1}{2}\left(I_4-1\right)^2
	+\frac{\gamma_2}{2}\left(I_5-1\right)^2+\frac{\xi_1}{2}\left(I_6-1\right)^2+\frac{\xi_2}{2}\left(I_7-1\right)^{\beta}\,.
\end{equation}
Then the energy per unit length along a single screw dislocation line is finite if $\beta<1$. To see this, we need to show that $2\pi\int_{0}^{R_o}\frac{\xi_2}{2}\left(I_7(\zeta)-1\right)^{\beta}\zeta d\zeta<\infty$ as the finiteness of the contribution of the other terms in the energy per unit length is trivial (cf. \eqref{inv-or-dis}). Noting that for a single screw dislocation with Burgers vector $b_i$, $b(R)=2\pi b_i\,\delta^2(R)$, and hence, $f(R)=\frac{b_i}{2\pi}H(R)$, we have
\begin{equation}
	\pi\xi_2\int_{0}^{R_o}\left(I_7(\zeta)-1\right)^{\beta}\zeta d\zeta=\pi\xi_2\int_{0}^{R_o}\Big[\frac{\alpha^4}{{\zeta}^2}(\zeta^2+f(\zeta)^2)-1\Big]^{\beta}\zeta d\zeta=\pi\xi_2\int_{0}^{R_o}\Big[\alpha^4-1+\frac{\alpha^4b_i^2}{4\pi^2\zeta^2}\Big]^{\beta}\zeta d\zeta<\infty\,,
\end{equation}
provided that $\beta<1$. Similarly, one can show that if the resultant longitudinal force, i.e., $F_Z=2\pi\int_{0}^{R_o}\hat{\sigma}^{zz}(\zeta)\zeta d\zeta$, induced by a single screw dislocation is finite for the isotropic base material with the energy function $W_{\mathrm{fib}}(I_1,I_2)$, so is the axial force for the fiber-reinforced material with the energy function \eqref{rem-ren} when $\beta<1$.  
\end{remark}




\subsection{A Cylindrically-Symmetric Distribution of Parallel Screw Disclocations in a Monoclinic Medium}
\label{scr-mo}

In the previous section, we assumed that the dislocated body is orthotropic in the reference configuration.
Instead, let us assume that the medium with the cylindrically-symmetric distribution of parallel screw dislocations is orthotropic in its current configuration such that the orthotropic axes are in the radial, circumferential, and axial directions in the ambient space. 
In the reference configuration, the material will be monoclinic such that $\mathbf{N}_3=\hat{\mathbf{R}}$ is perpendicular to the plane of $\mathbf{N}_1=\hat{\mathbf{\Theta}}$ and $\mathbf{N}_2=\hat{\mathbf{Z}}$.\footnote{Note that $\mathbf{N}_1$ and $\mathbf{N}_2$ are not orthogonal in the nontrivial geometry of the reference configuration.}
We assume the same class of deformations as was assumed in the previous section, and thus, $r(R)=\frac{1}{\sqrt{\alpha}}R$. Hence, the right Cauchy-Green deformation tensor is given by \eqref{disco-c}. 
The invariants of the strain energy function for the monoclinic material are given as
\begin{equation}
\begin{aligned}
I_1=&\operatorname{tr}(\mathbf{C})=\frac{2}{\alpha}+\frac{\alpha^2}{R^2}(R^2+f(R)^2)\,, \quad
I_2=
\frac{1}{2}\left[\operatorname{tr}(\mathbf{C}^2)-(\operatorname{tr}\mathbf{C})^2\right]
=\frac{1}{\alpha^2}+2\alpha+\alpha\frac{f(R)^2}{R^2}\,,\\	
I_4=&\frac{R^2}{\alpha(R^2+f(R)^2)}\,,~
I_5=\frac{R^2}{\alpha^2(R^2+f(R)^2)}\,,~
I_6=\alpha^2\,,~
I_7= \frac{\alpha^4}{R^2}(R^2+f(R)^2) \,,~ I_8=0\,.
\end{aligned}
\end{equation}
From \eqref{cauchymon}, the non-zero components of the Cauchy stress tensor read
\begin{align}
\sigma^{rr}=&\frac{2W_{I_2}}{\alpha ^2} \Big(\alpha
^3+\frac{\alpha ^3 f(R)^2}{R^2}+1\Big)+\frac{2 W_{I_1}}{\alpha }-p(R)\,,\label{sirr}  \\
\sigma^{\theta\theta}=&\frac{2 \alpha  W_{I_1}+2 \left(\alpha^3+1\right) W_{I_2}-\alpha^2p(R)}{\alpha  R^2}+\frac{2(\alpha W_{I_4}+2W_{I_5})}{\alpha(R^2+f(R)^2)}\,,\label{siph}\\
\sigma^{zz}=&\frac{2 \alpha}{R^2}  \Big[ \left(f(R)^2+R^2\right) (\alpha W_{I_1}+W_{I_2}+2 \alpha^3W_{I_7})+R^2 (W_{I_2}+\alpha   W_{I_6})\Big]-p(R)\,,\\
\sigma^{\theta z}=&-\frac{2 f(R)}{R^2} \left(\alpha  W_{I_1}+W_{I_2}+\alpha ^3 W_{I_7}\right)-\frac{2f(R)}{R^2+f(R)^2}W_{I_5}+\frac{\alpha}{(R^2+f(R)^2)^{\frac{1}{2}}}W_{I_8}\,.
\label{teta}
\end{align}
Note that for the stress to vanish when $\alpha=1$ and the body is dislocation-free, i.e., $f(R)=0$ (identically), one needs to have $\left(W_{I_4}+2W_{I_5}\right)=\left(W_{I_6}+2W_{I_7}\right)=W_{I_8}=0$, evaluated at $I_1=I_2=3$, $I_4=I_5=I_6=I_7=1$, $I_8=0$. The equilibrium equation implies that $p'(R)=S(R)$, where
\begin{equation}
\begin{aligned}
	S(R)=&-\frac{1}{\alpha ^4 R^3}\bigg[-4 \alpha f(R)\left(Rf'(R)-f(R)\right) \left(\alpha ^4 W_{I_1I_1}
	+\alpha ^3W_{I_1I_2}+\alpha ^6 W_{I_1I_7}-\frac{R^4(\alpha W_{I_1I_4}
	+W_{I_1I_5})}{\left(f(R)^2+R^2\right)^2}\right) \\
	&-4 f(R) \Big(\alpha ^3+\frac{\alpha ^3f(R)^2}{R^2}+1\Big) \left(R f'(R)-f(R)\right) 
	\bigg\{\alpha^4 W_{I_1I_2}+\alpha ^3W_{I_2I_2}+\alpha ^6 W_{I_2I_7} \\
	& -\frac{  R^4(\alpha W_{I_2I_4}+W_{I_2I_5})}{\left(f(R)^2+R^2\right)^2}\bigg\}
	+4 \alpha ^5 f(R) W_{I_2} \left(f(R)-Rf'(R)\right) 
	+\frac{2\alpha ^2 R^4 (\alpha  W_{I_4}+2W_{I_5})}{f(R)^2+R^2} \\
	&-2\alpha^5 f(R)^2 W_{I_2}\bigg]\,.
\end{aligned}
\end{equation}
Assuming that the surface $R=R_o$ is traction-free the pressure is obtained as
\begin{equation}
p(R)=\int_{R_o}^{R}S(\zeta)d\zeta+\frac{2}{\alpha ^2}  \Big(\alpha
^3+\frac{\alpha ^3 f(R_o)^2}{R_o^2}+1\Big)W_{I_2}|_{R=R_o}+\frac{2}{\alpha } W_{I_1}|_{R=R_o}\,.
\end{equation}
Let us consider the following model for the strain energy function
\begin{equation}
W=W\left(I_1,I_2,I_4,I_5,I_6,I_7,I_8\right)=W_{\mathrm{iso}}\left(I_1,I_2\right)+W_{\mathrm{fib}}^{\Theta}\left(I_4,I_5\right)+W_{\mathrm{fib}}^Z\left(I_6,I_7\right)+W_{\mathrm{fib}}^{Z\Theta}(I_8)\,,
\label{reimoe}
\end{equation}
where $W_{\mathrm{iso}}$ describes that part of the energy function pertaining to the isotropic base material, while $W_{\mathrm{fib}}^{\Theta}$ and $W_{\mathrm{fib}}^{Z}$ represent the reinforcement effects in the circumferential and axial directions. $W_{\mathrm{fib}}^{Z\Theta}(I_8)$ models the coupling between the axial and circumferential fibers. Note, however, that $I_8=0$, and for the stress to vanish for the dislocation-free body, one needs $W_{I_8}=0$ at $I_8=0$, which implies that $W_{\mathrm{fib}}^{Z\Theta}(I_8)=0$, i.e., the coupling term must vanish. A way out would be to require that the coupling term depend on some other invariants as well, e.g., one can define $W_{\mathrm{fib}}^{Z\Theta}(I_1,I_8)=\eta I_8(I_1-3)$ for some positive constant $\eta$.
For the sake of simplicity, as an example, we consider a fiber-reinforced Mooney-Rivlin material with the following energy function
\begin{equation}
\begin{aligned}
W(I_1,I_2,I_4,I_5,I_6,I_7) =&\frac{\mu_1}{2}\left(I_1-3\right)+\frac{\mu_2}{2}\left(I_2-3\right)+\frac{\lambda_1}{2}\left(I_4-1\right)^2\\
&+\frac{\lambda_2}{2}\left(I_5-1\right)^2+\frac{\xi_1}{2}\left(I_6-1\right)^2+\frac{\xi_2}{2}\left(I_7-1\right)^2\,.
\label{sten1}
\end{aligned}
\end{equation} 
Therefore, one obtains
\begin{equation}\label{s-mo}
\begin{split}
S(R)=&\frac{1}{\alpha ^2 R^3 \left(f(R)^2+R^2\right)}\Big[ \alpha ^3 \mu_2  f(R) (R^2+f(R)^2)(2Rf'(R)-f(R))\\&+2 R^4 \Big\{\alpha 
\lambda_1-\frac{
R^2}{f(R)^2+R^2}(\frac{2 \lambda_2}{\alpha^2}+\lambda_1)+2 \lambda_2\Big\}\Big]\,.
\end{split}
\end{equation}
Thus, the physical components of the stress are given as
\begin{align}
\hat{\sigma}^{rr}&=\alpha\mu_2\left( \frac{  f(R)^2}{R^2}-\frac{f(R_o)^2}{R_o^2}\right)+\int_{R}^{R_o}S(\zeta)d\zeta\,,\\
\begin{split}
\hat{\sigma}^{\theta\theta}&=\int_{R}^{R_o}S(\zeta)d\zeta-\alpha\mu_2\frac{f(R_o)^2}{R_o^2}+\frac{2R^2}{\alpha^2(R^2+f(R)^2)}\Big[\alpha\lambda_1\Big(\frac{R^2}{\alpha(R^2+f(R)^2)}-1\Big)\\&+2\lambda_2\Big(\frac{R^2}{\alpha^2(R^2+f(R)^2)}-1\Big)\Big]\,,
\end{split}\\
\begin{split}
\hat{\sigma}^{zz}&=2\alpha^2\xi_1(\alpha^2-1)+4\alpha^4\xi_2\Big(1+\frac{f(R)^2}{R^2}\Big)\Big[\alpha^4\Big(1+\frac{f(R)^2}{R^2}\Big)-1\Big]+\int_{R}^{R_o}S(\zeta)d\zeta\\&+\alpha\mu_2\left( \frac{  f(R)^2}{R^2}-\frac{f(R_o)^2}{R_o^2}\right)+(\alpha\mu_1+\mu_2)(\alpha-\frac{1}{\alpha^2})+\alpha^2\mu_1\frac{f(R)^2}{R^2}\,,
\end{split}\\
\begin{split}
\hat{\sigma}^{\theta z}&=-\frac{1}{\sqrt{\alpha}}\frac{f(R)}{R}(\alpha\mu_1+\mu_2)-2\xi_2\alpha^{\frac{5}{2}}\frac{f(R)}{R}\Big[\alpha^4\Big(1+\frac{f(R)^2}{R^2}\Big)-1\Big]\\&-\frac{2\lambda_2}{\sqrt{\alpha}}\frac{Rf(R)}{R^2+f(R)^2}\Big(\frac{R^2}{\alpha^2(R^2+f(R)^2)}-1\Big)\,.
\end{split}
\end{align}

\begin{figure}
	\centering
	\begin{center}
		\makebox[0pt]{\includegraphics[trim = 14.0cm 0.4cm 14.5cm 0.7cm, clip, scale=0.35]{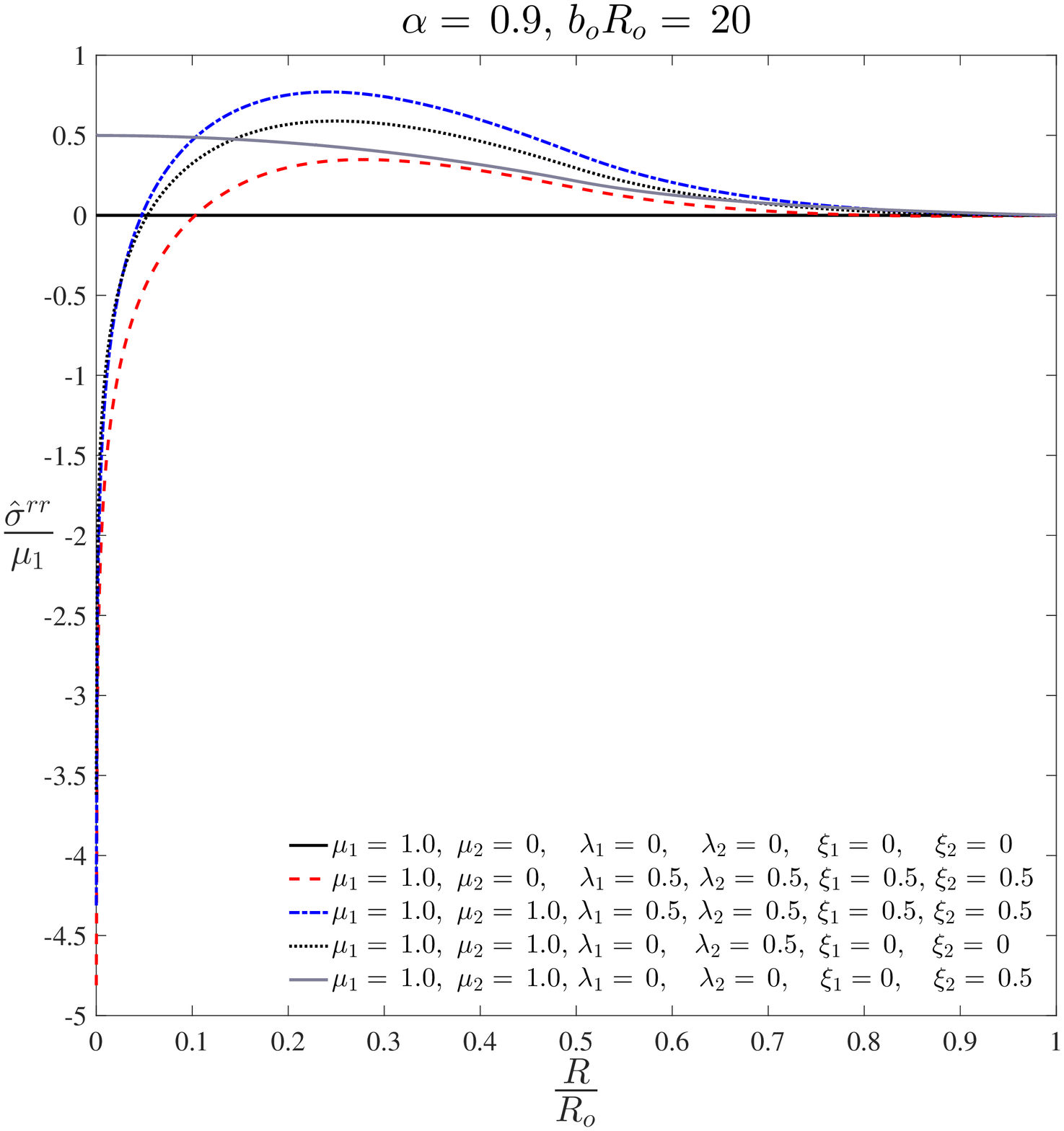}
			\includegraphics[trim = 14.0cm 0.4cm 14.5cm 0.7cm, clip, scale=0.35]{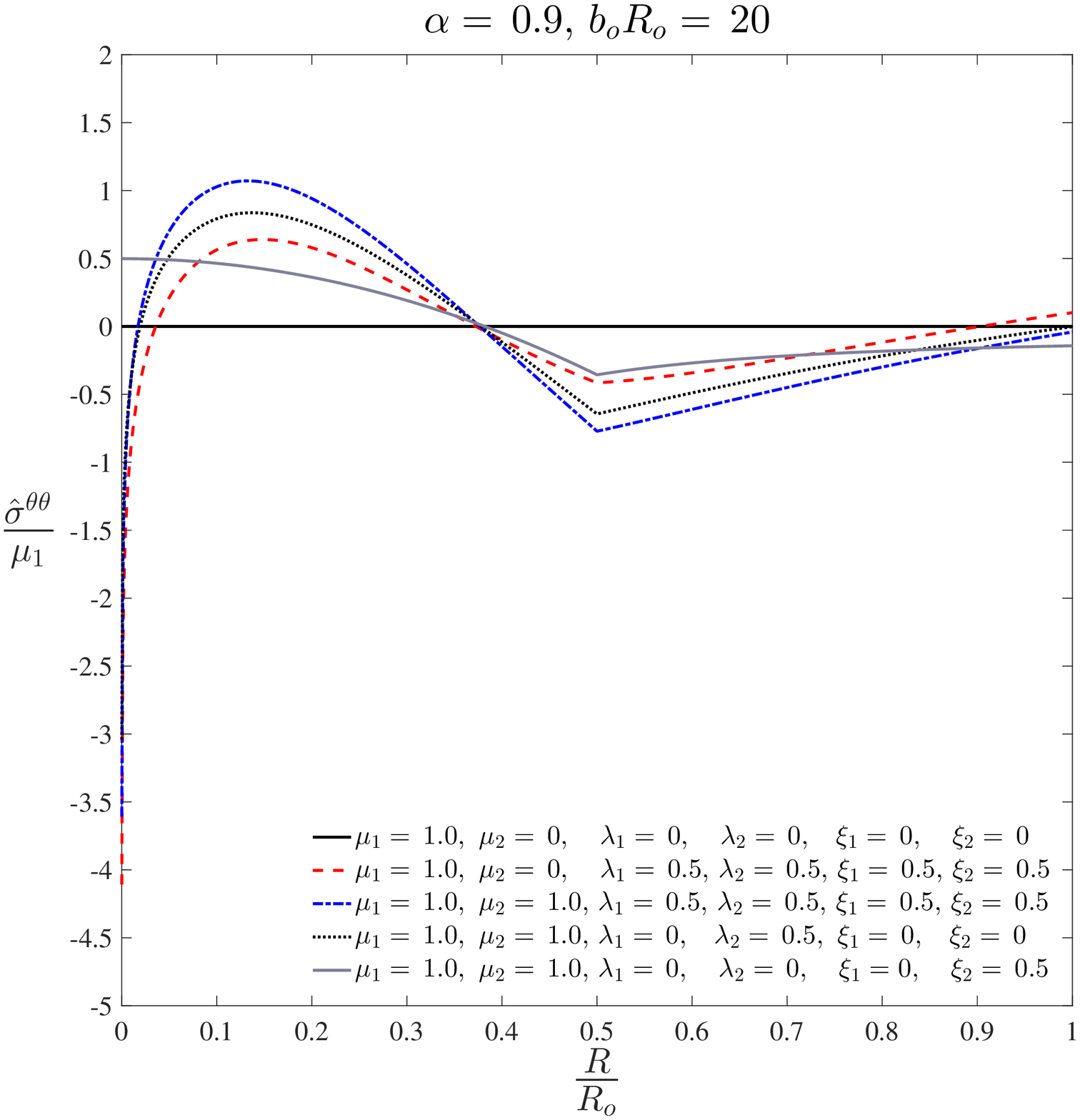}}
	\end{center}
	\begin{center}
		\makebox[0pt]{\includegraphics[trim = 14.0cm 0.4cm 14.5cm 0.7cm, clip, scale=0.35]{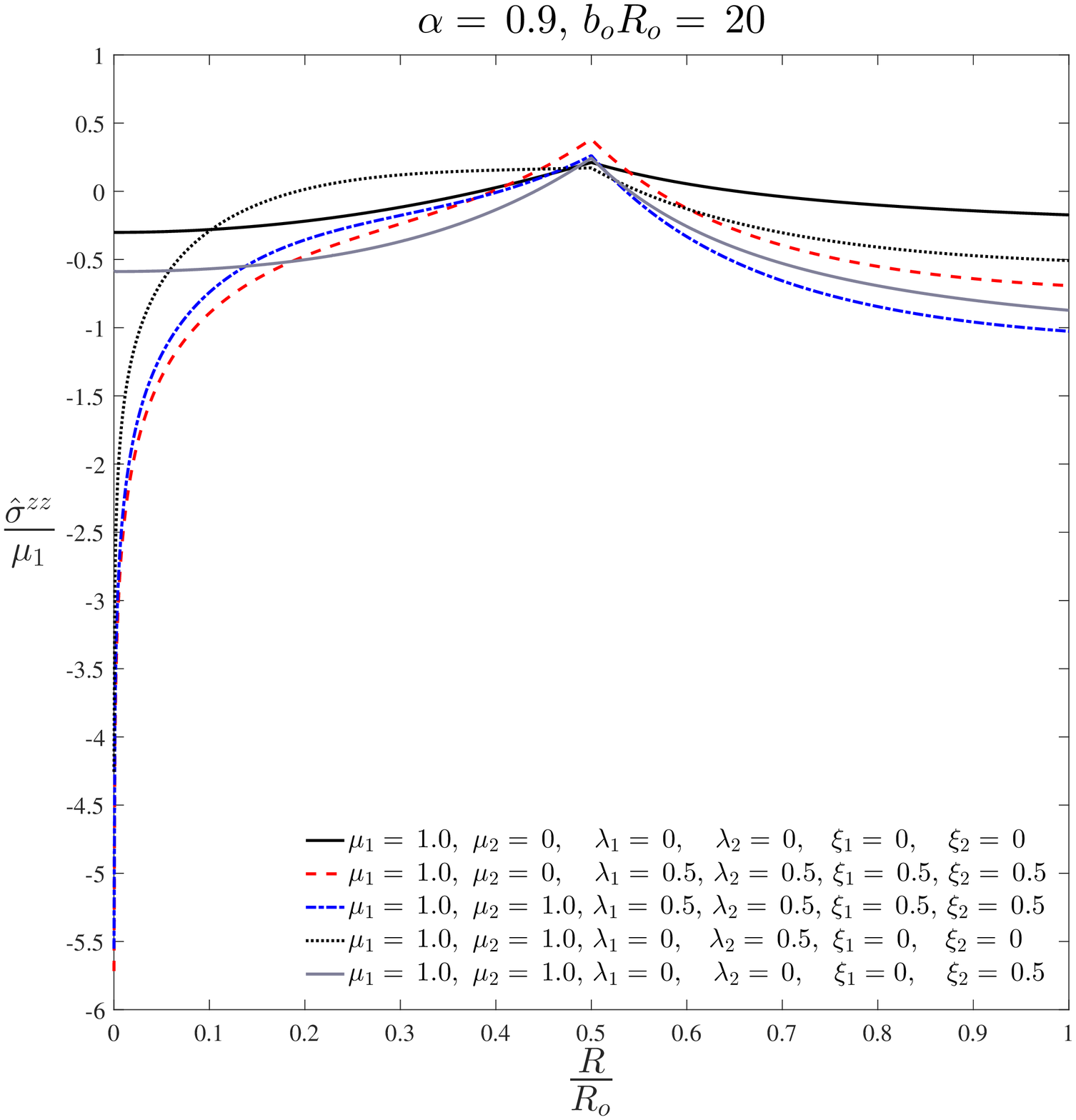}
			\includegraphics[trim = 14.0cm 0.4cm 14.5cm 0.7cm, clip, scale=0.35]{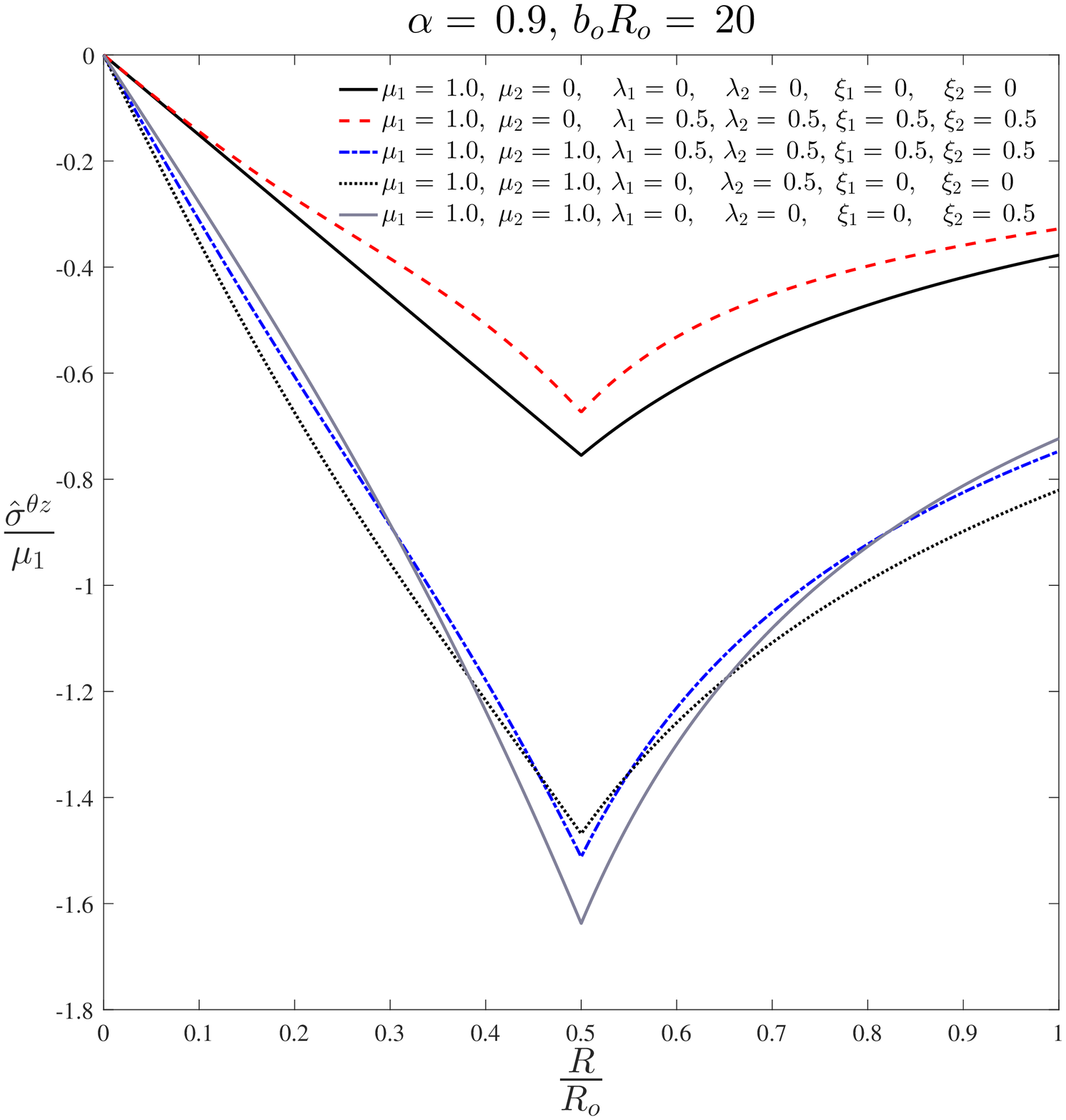}}
	\end{center}
	\vspace*{-0.2in}
	\caption{Stress distribution in a medium with the constitutive equation \eqref{sten1} and the dislocation distribution $\eqref{uni-bur}$ such that $R_i/R_o=0.5$, $b_oR_o=20$, and $\alpha=0.9$ for different values of the constitutive parameters.} 
	\label{sa}
\end{figure}

\begin{remark}
For an arbitrary cylindrically-symmetric distribution of parallel screw dislocations with a smooth Burgers' vector density $b(R)$ in a monoclinic material, the pressure, and hence, $\hat{\sigma}^{rr}$, $\hat{\sigma}^{\theta\theta}$, and $\hat{\sigma}^{zz}$ have a logarithmic singularity on the dislocation axis unless $\alpha=1$. Nevertheless, the shear component $\hat{\sigma}^{\theta z}$ is finite and vanishes at $R=0$. This is because as $R\to 0$, we have $f(R)=\frac{b(0)}{4\pi}R^2+\mathcal{O}(R^3)$, and thus, from \eqref{s-mo}
\begin{equation}
 S(R)=\frac{2\left(\alpha-1\right)}{\alpha^2}\left[\lambda_1+\frac{2\lambda_2}{\alpha}\left(1+\frac{1}{\alpha}\right)\right]\frac{1}{R}+\mathcal{O}(R)\,.
\end{equation}
Therefore, $p(R)=C-\frac{2\left(\alpha-1\right)}{\alpha^2}\left[\lambda_1+\frac{2\lambda_2}{\alpha}\left(1+\frac{1}{\alpha}\right)\right]\ln\frac{R}{R_o}+\mathcal{O}(R^2)$ as $R\to 0$, where $C$ is a constant. It is easy to see that when $\alpha=1$, the stress is finite and $\hat{\sigma}^{rr}=\hat{\sigma}^{\theta\theta}=\hat{\sigma}^{zz}$ at $R=0$.
\end{remark}
In Fig.~\ref{sa} the stress field is shown for the dislocation distribution \eqref{uni-bur}, where $R_i/R_o=0.5$ and $b_0R_o=20$ for different values of the constitutive parameters given by \eqref{sten1}.




\subsection{A Parallel Cylindrically Symmetric Distribution of Wedge Disclinations in an Orthotropic Medium}
\label{wed-or}

Let us consider a parallel cylindrically-symmetric distribution of wedge disclinations in an infinite orthotropic medium in the reference configuration. 
In the cylindrical coordinates $\left(R,\Theta,Z\right)$, assume that the material orthotropic axes are in the $R$, $\Theta$, and $Z$ directions. The radial density of the wedge disclinations is denoted by $w(R)$. The material manifold for a body having a distribution of wedge disclinations is a Riemannian manifold with a non-vanishing curvature. The material metric for the disclinated body is given by \citep{yavari2013riemann} 
\begin{equation}
\label{_G2}
\mathbf{G}=
\left(
\begin{array}{ccc}
1&0&0\\
0&f(R)^2&0\\
0&0&1
\end{array}
\right)\,,
\end{equation}
where $f''(R)=-\frac{R}{2\pi}w(R)$. 
The ambient space is endowed with the Euclidean metric $\mathbf{g}=\operatorname{diag}\{1,r^2,1\}$. We embed the material manifold into the ambient space by looking for mappings\footnote{Note that for the class of deformations that is considered, the material will be orthotropic in its current configuration as well. The orthotropic axes in the current configuration will be in the radial, circumferential, and axial directions (similar to those in the reference configuration).} of the form $\left(r,\theta,z\right)=\left(r(R),\Theta,\alpha Z\right)$, where $\alpha$ is a constant representing the axial stretch of the bar that depends on the axial boundary conditions. Therefore, the deformation gradient reads $\mathbf{F}=\mathrm{diag}\left(r'(R),1,\alpha\right)$. Incompressibility constraint dictates that $J=\sqrt{\frac{\mathrm{det} \mathbf g}{\mathrm{det} \mathbf G}} \mathrm{det} \mathbf F=\alpha\frac{r(R)}{f(R)}r'(R)=1$.
Thus, imposing $r(0)=0$, we have $r(R)=\left(\frac{2}{\alpha}\int_{0}^{R}f(\xi)d\xi\right)^{\frac{1}{2}}$. The right Cauchy-Green deformation tensor reads $\mathbf{C}=\operatorname{diag}\left\{\frac{1}{\alpha^2}\frac{f(R)^2}{r(R)^2},\frac{r(R)^2}{f(R)^2},\alpha^2\right\}$. From \eqref{invaror}, the invariants of the strain energy function are simplified to read
\begin{equation}
\begin{aligned}
I_1=&\operatorname{tr}(\mathbf{C})=\alpha^2+\frac{1}{\alpha^2}\frac{f(R)^2}{r(R)^2}+\frac{r(R)^2}{f(R)^2}\,, \quad
I_2=
\frac{1}{2}(\operatorname{tr}(\mathbf{C}^2)-\operatorname{tr}(\mathbf{C})^2)=\frac{1}{\alpha^2}+\alpha^2\frac{r(R)^2}{f(R)^2}+\frac{f(R)^2}{r(R)^2}\,,\\	
I_4=&\frac{1}{\alpha^2}\frac{f(R)^2}{ r(R)^2}\,,\qquad
I_5=\frac{1}{\alpha^4}\frac{f(R)^4}{ r(R)^4}\,,\qquad
I_6=\alpha^2\,, \quad
I_7= \alpha^4 \,.
\end{aligned}
\end{equation}
The non-zero physical components of the Cauchy stress are as follows\footnote{When the body is disclination-free $f(R)=R$, and the stress vanishes if the energy function satisfies \eqref{compor1}.} 
\begin{align}
\hat{\sigma}^{rr}=&\frac{2}{\alpha^2}\frac{f(R)^2}{r(R)^2}\left(W_{I_1}+\alpha^2W_{I_2}+W_{I_4}\right)+\frac{4}{\alpha^4}\frac{f(R)^4}{r(R)^4}W_{I_5}+\frac{2}{\alpha^2}W_{I_2}-p(R)\,,\\
\hat{\sigma}^{\theta\theta}=&2\frac{r(R)^2}{f(R)^2}\left(W_{I_1}+\alpha^2W_{I_2}\right)+\frac{2}{\alpha^2}W_{I_2}-p(R)\,,\\
\hat{\sigma}^{zz}=&2\alpha^2\left(W_{I_1}+W_{I_6}+2\alpha^2W_{I_7}\right)+2W_{I_2}\left(\frac{f(R)^2}{r(R)^2}+\alpha^2\frac{r(R)^2}{f(R)^2}\right)-p(R)\,.
\end{align}
The equilibrium equation implies that $p'(R)=k(R)$, where
\begin{equation}
\begin{aligned}
	k(R) &=\frac{2}{\alpha ^8 f^3 r^9} \bigg[2 \alpha ^4 f^4 r^7 f' \left(\alpha ^2
	W_{I_1}+W_{I_1I_2}-2 W_{I_1I_5}+\alpha ^4 W_{I_2}
	+\alpha^2 W_{I_2I_2}+W_{I_2I_4}-2 \alpha ^2 W_{I_2I_5}+\alpha ^2	W_{I_4}\right)\\
	&+2 \alpha ^2 f^6 r^5 f' \left(\alpha ^2W_{I_1I_1}+2 \alpha ^4 W_{I_1I_2}+2 \alpha ^2 W_{I_1I_4}
	+\alpha^6 W_{I_2I_2}+2 \alpha ^4 W_{I_2I_4}+2 W_{I_2I_5}+\alpha ^2W_{I_4I_4}
	+4 \alpha ^2 W_{I_5}\right)\\
	&-2 \alpha ^6 f^2 r^9 f'\left(W_{I_1I_1}+2 \alpha ^2 W_{I_1I_2}+W_{I_1I_4}
	+\alpha ^4W_{I_2I_2}+\alpha ^2 W_{I_2I_4}\right)\\
	&+8 \alpha ^2 f^8 r^3f' \left(W_{I_1I_5}+\alpha ^2 W_{I_2I_5}+W_{I_4I_5}\right)
	+8f^{10} W_{I_5I_5} r f'-2 \alpha ^6 r^{11} f'\left(W_{I_1I_2}+\alpha ^2 W_{I_2I_2}\right)\\
	&-\alpha ^3 f^6 r^5 \left(\alpha ^2W_{I_1}+2 W_{I_1I_2}-4 W_{I_1I_5}+\alpha ^4 W_{I_2}
	+2\alpha ^2 W_{I_2I_2}+2 W_{I_2I_4}-4 \alpha ^2 W_{I_2I_5}
	+\alpha^2 W_{I_4}\right)\\&-\alpha ^5 f^2 r^9 \left(\alpha ^2 W_{I_1}
	-2W_{I_1I_2}+\alpha ^4 W_{I_2}-2 \alpha ^2 W_{I_2I_2}\right)-8 \alpha  f^{10} r \left(W_{I_1I_5}
	+\alpha ^2W_{I_2I_5}+W_{I_4I_5}\right)\\
	&-2\alpha  f^8 r^3 \left(\alpha ^2 W_{I_1I_1}+2 \alpha ^4W_{I_1I_2}+2 \alpha ^2 W_{I_1I_4}
	+\alpha ^6 W_{I_2I_2}+2 \alpha^4 W_{I_2I_4}+2 W_{I_2I_5}+\alpha ^2 W_{I_4I_4}
	+3 \alpha ^2W_{I_5}\right)\\&+2 \alpha ^5 f^4 r^7 \left(W_{I_1I_1}
	+2 \alpha ^2W_{I_1I_2}+W_{I_1I_4}+\alpha ^4 W_{I_2I_2}+\alpha ^2W_{I_2I_4}\right)
	-\frac{8 f^{12} W_{I_5I_5}}{\alpha r}\bigg]\,.
\end{aligned}
\end{equation}
Assuming \eqref{sten0} for the energy function, one obtains
\begin{equation}\label{pr-disc}
\begin{aligned}
p'(R)=&-\frac{1}{\alpha ^9 f r^{10}}\Big[-2 \alpha ^7 f^2 r^8 f' \left(\alpha ^2 \mu_2-2
	\gamma_1+\mu_1\right)+\alpha ^5 f^4 r^6 \left\{\alpha 
	\left(\alpha ^2 \mu_2-2 \gamma_1+\mu_1\right)-8
	(\gamma_1-2 \gamma_2) f'\right\}\\&-32 \alpha  \gamma
	_2 f^8 r^2 f'+6 \alpha ^4 (\gamma_1-2 \gamma_2)
	f^6 r^4+28 \gamma_2 f^{10}+\alpha ^8 r^{10} \left(\alpha ^2
	\mu_2+\mu_1\right)\Big]\,.
\end{aligned}
\end{equation}
Knowing that the traction vanishes on the outer boundary $R=R_o$, one finds
\begin{equation}
	p(R_o)=\frac{1}{\alpha^2}\frac{f(R_o)^2}{r(R_o)^2}
	\Big[\mu_1+\alpha^2\mu_2+2\gamma_1\Big(\frac{1}{\alpha^2}\frac{f(R_o)^2}{r(R_o)^2}-1\Big)\Big]
	+\frac{4\gamma_2}{\alpha^4}\frac{f(R_o)^4}{r(R_o)^4}\left(\frac{1}{\alpha^4}\frac{f(R_o)^4}{ r(R_o)^4}-1\right)
	+\frac{\mu_2}{\alpha^2}\,.
\end{equation}
Therefore, $p(R)=\int_{R_o}^{R}p'(\xi)d\xi+p(R_o)$. The stress components are simplified and read
\begin{align}
	\hat{\sigma}^{rr}=&\frac{1}{\alpha^2}\frac{f(R)^2}{r(R)^2}
	\Big[\mu_1+\alpha^2\mu_2+2\gamma_1\Big(\frac{1}{\alpha^2}\frac{f(R)^2}{r(R)^2}-1\Big)\Big]
	+\frac{4\gamma_2}{\alpha^4}\frac{f(R)^4}{r(R)^4}\left(\frac{1}{\alpha^4}\frac{f(R)^4}{ r(R)^4}-1\right)
	+\frac{\mu_2}{\alpha^2}-p(R)\,,\\
	\hat{\sigma}^{\theta\theta}=&\frac{r(R)^2}{f(R)^2}\left(\mu_1+\alpha^2\mu_2\right)+\frac{\mu_2}{\alpha^2}-p(R)\,,\\
	\hat{\sigma}^{zz}=&\alpha^2\left[\mu_1+2\xi_1(\alpha^2-1)+4\alpha^2\xi_2(\alpha^4-1)\right]
	+\mu_2\left(\frac{f(R)^2}{r(R)^2}+\alpha^2\frac{r(R)^2}{f(R)^2}\right)-p(R)\,.
\end{align}

\begin{example}
For a uniform disclination  distribution $w(R)=w_o$, one has $f''(R)=-\frac{R}{2\pi}w_o$, and thus, $f(R)=R-\frac{w_o}{12\pi}R^3$. Therefore
	\begin{equation}
	r(R)=\frac{R}{\alpha^{\frac{1}{2}}}\Big(1-\frac{w_o}{24\pi}R^2\Big)^{\frac{1}{2}}\,,
	\end{equation} 
provided that $w_o<24\pi/R_o^2$. 

\begin{remark}
For the uniform disclination distribution, the stress field exhibits a logarithmic singularity on the disclinations axis unless the axial stretch $\alpha=1$. Moreover, when $\alpha=1$, the stress is finite and hydrostatic at $R=0$. To see this, as $R\to 0$, we have $r(R)=\frac{R}{\alpha^{\frac{1}{2}}}+\mathcal{O}(R^3)$, and $f(R)=R+\mathcal{O}(R^3)$. From \eqref{pr-disc}, therefore
\begin{equation}
p(R)=C+\frac{2(1-\alpha)}{\alpha^4 } \left[\alpha ^2 \gamma_1+2 \gamma_2(1+\alpha)\right]\ln{R}+\mathcal{O}(R^2)\,,
\end{equation}
where $C$ is a constant. Hence, the stress is logarithmically unbounded at $R=0$ unless $\alpha=1$. Similar to the case of parallel screw dislocations in an orthotropic medium (cf. Remark.~\ref{sin-sc}), the singularity arises as a result of radial reinforcement effects, and does not, in particular, occur in isotropic materials. Note that for $\alpha=1$, at $R=0$, one has $\hat{\sigma}^{rr}=\hat{\sigma}^{\theta\theta}=\hat{\sigma}^{zz}=\mu_1+2\mu_2+\int_{0}^{R_o}p'(\xi)d\xi-p(R_o)$.
\end{remark}

In Fig.~\ref{sirr-disc}, we show the variation of the stress components for the uniform disclination distribution with $w_o=8\pi/R_o^2$ and for some different values of the constitutive parameters.\footnote{Note that the numerical values shown in  \citep{yavari2013riemann}'s Fig.~4 are not correct. This was caused by a typo in the sign of the integral term in the numerical evaluation of the pressure function from Eq. (4.23). In other words, the numerical values in that figure correspond to the following (incorrect) relation for the pressure with a positive sign for the integral term
\begin{equation*}
p(R)=\mu\frac{f^2(R_o)}{r^2(R_o)}+\mu\int_{R}^{R_o}\left[\frac{f(\eta)f'(\eta)}{\int_{0}^{\eta}f(\xi)d\xi}-\frac{f^3(\eta)}{4(\int_{0}^{\eta}f(\xi)d\xi)^2}-\frac{1}{f(\eta)}\right]d\eta\,.
\end{equation*}
} 
\begin{figure}
	\centering
	\begin{center}
    \makebox[0pt]{\includegraphics[trim = 13.9cm 0.2cm 14cm 0cm, clip, scale=0.35]{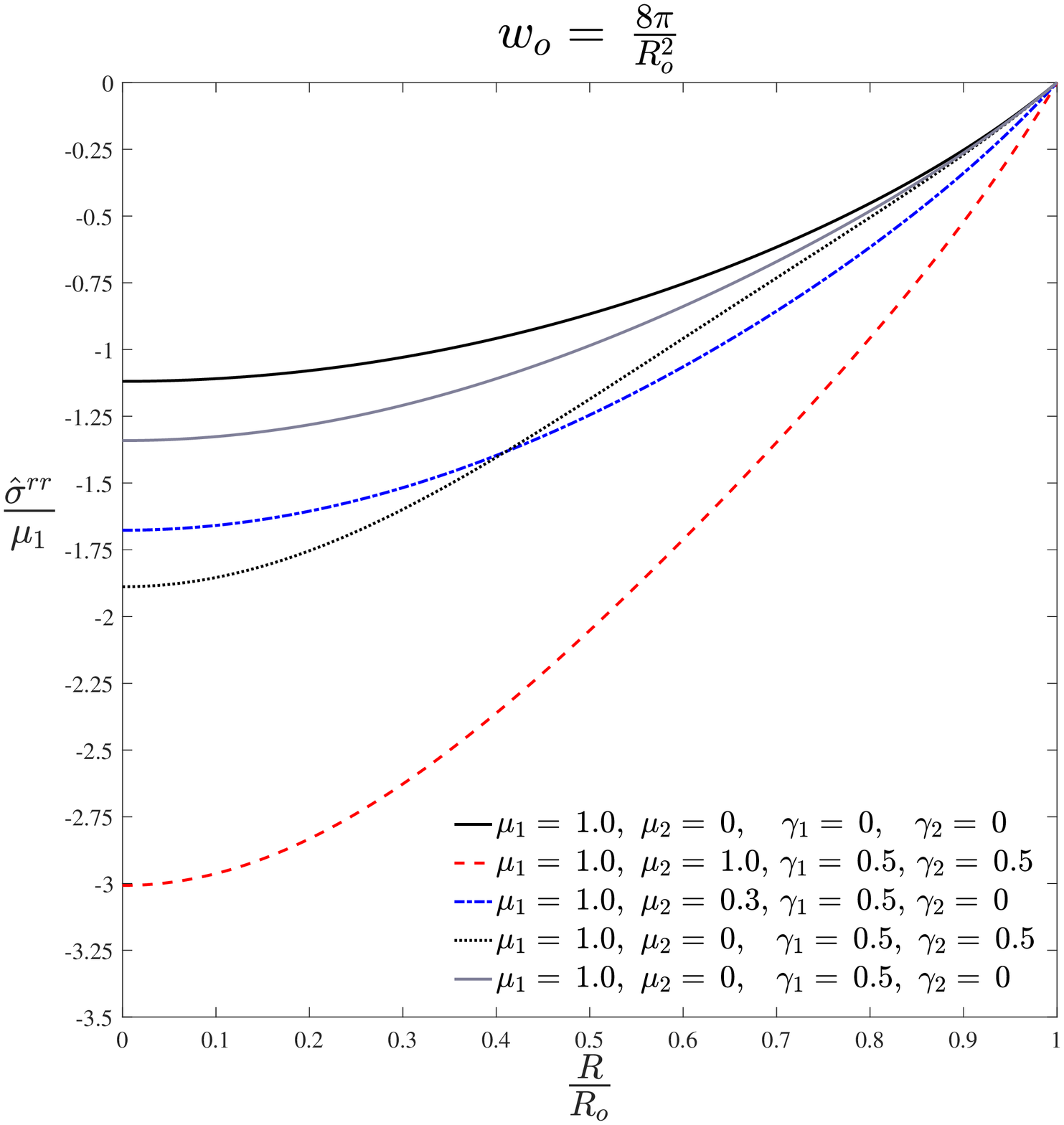}
    \includegraphics[trim = 14cm 0.2cm 14cm 0cm, clip, scale=0.35]{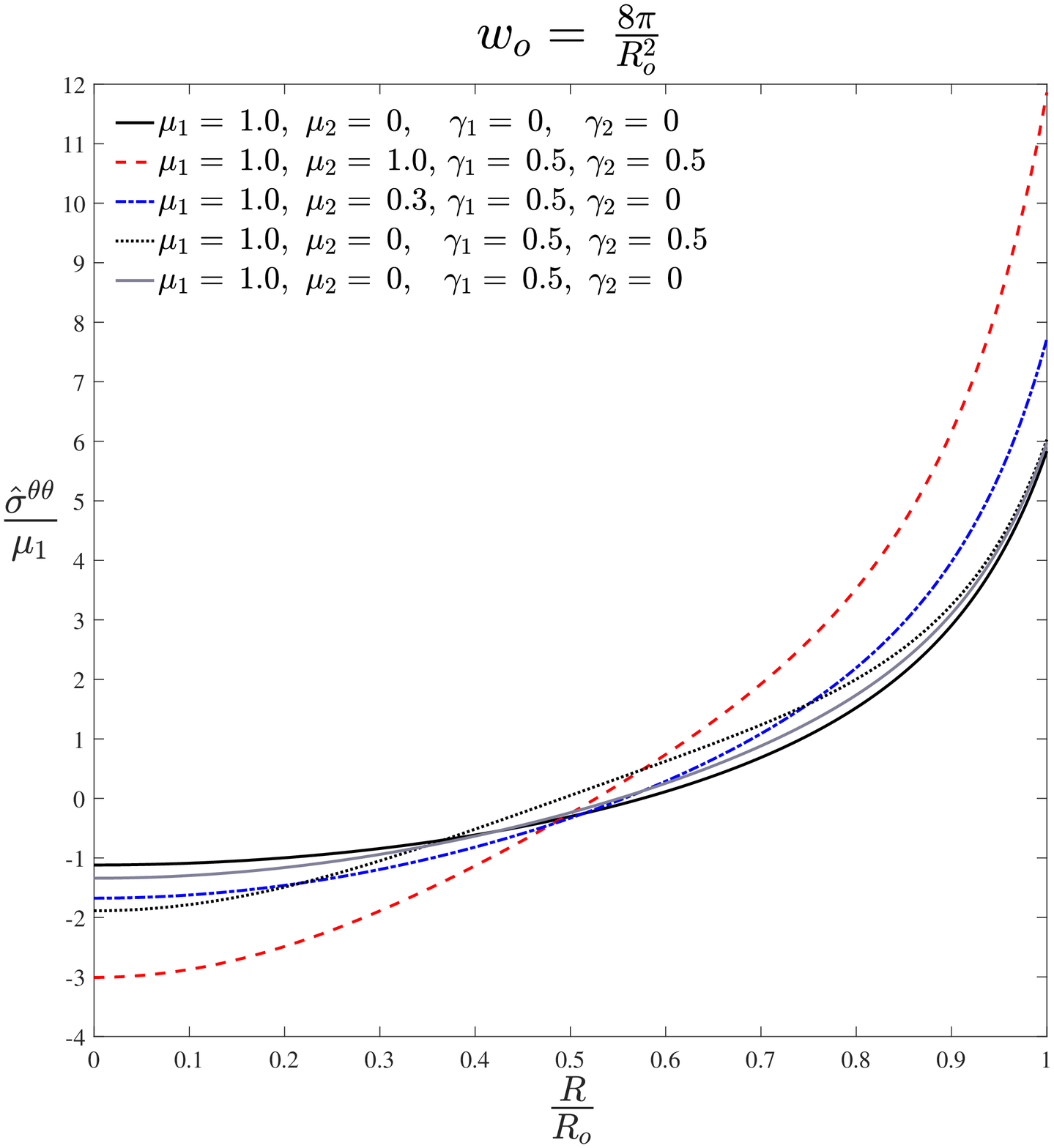}}
    \end{center}
	\vspace*{-0.18in}
	\caption{$\hat{\sigma}^{rr}$ and $\hat{\sigma}^{\theta\theta}$ distributions for different values of the constitutive parameters for a uniform disclination distribution with $\omega_o=8\pi/R_o^2$ such that $\alpha=1$.} 
	\label{sirr-disc}
\end{figure}
\end{example}

\begin{example}
For a single wedge disclination $\omega(R)=2\pi\Theta_o\delta^2(R)$, where $\Theta_o$ is the angle of the wedge shape region that is removed in Volterra's cut-and-weld operation (see \citep{yavari2013riemann} for more details). Therefore, $f''(R)=-\frac{\Theta_o}{2\pi}\delta (R)$, which implies that $f(R)=R(1-\frac{\Theta_o}{2\pi})$, and thus, $r(R)=\frac{R}{\alpha^{\frac{1}{2}}}(1-\frac{\Theta_o}{2\pi})^{\frac{1}{2}}$. Fig.~\ref{sirr-disc-sin} illustrates the stress distribution for different values of the reinforcement and the base material parameters in the case of a single wedge disclination of positive sign with $\Theta_o=\frac{\pi}{2}$.
\begin{figure}
	\centering
	\begin{center}
	\makebox[0pt]{\includegraphics[trim = 13.9cm 0.2cm 14cm 0cm, clip, scale=0.35]{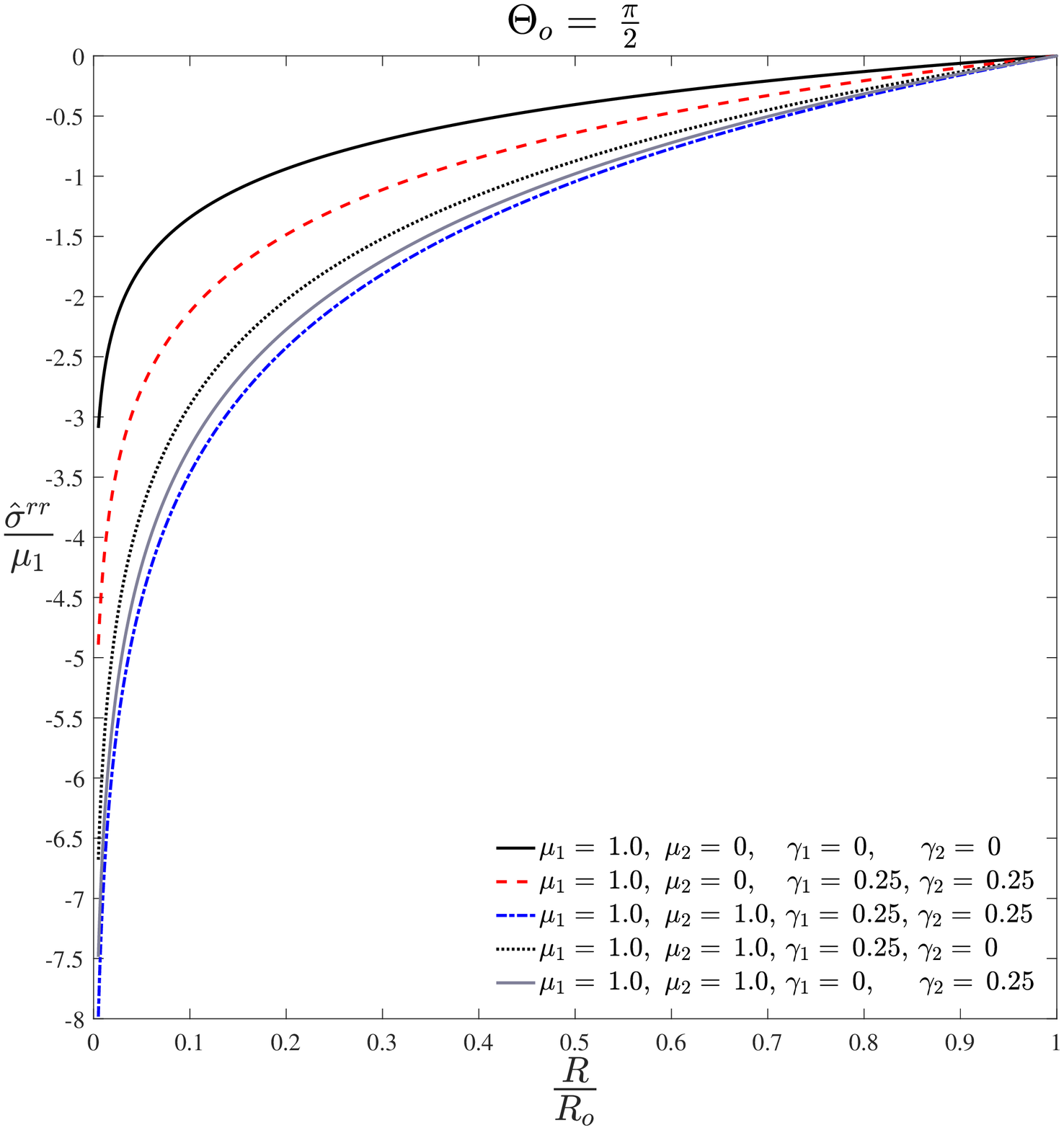}
	\includegraphics[trim = 14cm 0.2cm 14cm 0cm, clip, scale=0.35]{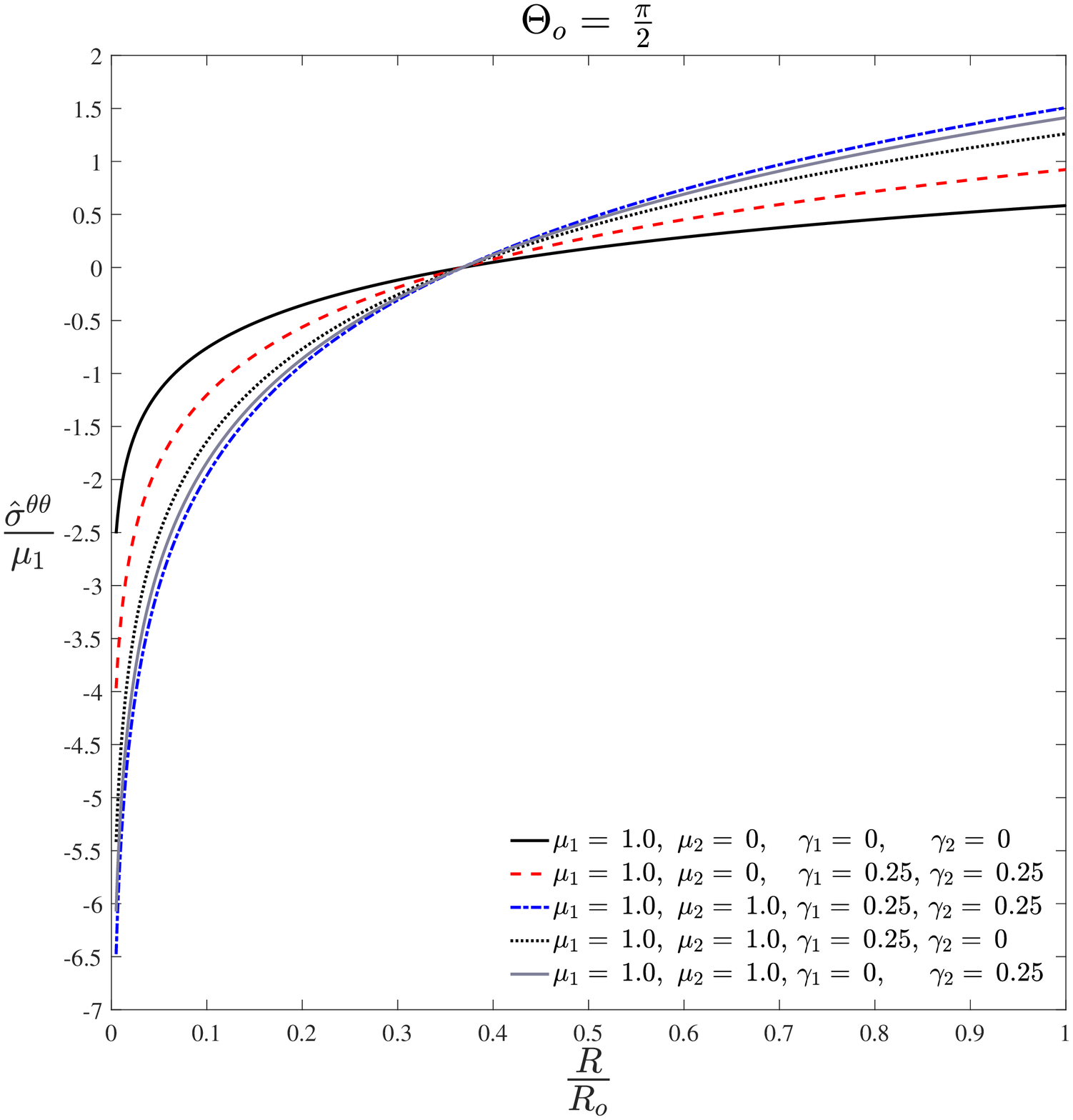}}
	\end{center}
	\vspace*{-0.18in}
	\caption{$\hat{\sigma}^{rr}$ and $\hat{\sigma}^{\theta\theta}$ distribution for different values of the constitutive parameters for a single positive wedge disclination with $\Theta_o=\pi/2$ such that $\alpha=1$.} 
	\label{sirr-disc-sin}
\end{figure}
\end{example}

\subsection{Distributed Edge Dislocations in an Orthotropic Medium}
\label{edge}

Next, we consider a distribution of edge dislocations in an orthotropic medium such that the material preferred directions are parallel to the Cartesian axes in the Cartesian coordinates $(X,Y,Z)$. Let us consider the orthonormal frame field $\{\mathbf{e}_{\alpha}(X,Y,Z)\}_{\alpha=1}^3$, where $\mathbf{e}_1$, $\mathbf{e}_2$, and $\mathbf{e}_3$ are in the $X$, $Y$, and $Z$-directions, respectively. We assume that the edge dislocation distribution consists of dislocations with i) the dislocation line parallel to the $Z$-axis such that the Burgers' vector density is given by $b_1(Z)\mathbf{e}_1+c_1(Z)\mathbf{e}_2$, ii) $X$-oriented Burgers' vector density $b_2(X,Y,Z)\mathbf{e}_1$ such that the dislocation line is parallel to the $Y$-axis, iii) $Y$-oriented Burgers' vector $c_2(X,Y,Z)\mathbf{e}_2$ with the dislocation line parallel to the $X$-axis. Let us consider the following co-frame field  
\begin{equation}\label{fr-fi}
\vartheta^1=e^{\xi(Z)+\gamma(Y)}dX\,,\quad \vartheta^2=e^{\eta(Z)+\lambda(X)}dY\,, \quad \vartheta^3=e^{\psi(Z)}dZ\,,
\end{equation}
where $\xi(Z)$, $\gamma(Y)$, $\eta(Z)$, $\lambda(X)$, and $\psi(Z)$ are scalar functions to be determined. The corresponding frame field reads
\begin{equation}
\mathbf{e}_1=e^{-\xi(Z)-\gamma(Y)}\partial_X\,,\quad \mathbf{e}_2=e^{-\eta(Z)-\lambda(X)}\partial_Y\,,\quad\mathbf{e}_3=e^{-\psi(Z)}\partial_Z\,.
\end{equation}
Note that $\mathbf{G}=\delta_{\alpha\beta}\vartheta^{\alpha}\otimes\vartheta^{\beta}$, and thus
\begin{equation}\label{G-edge}
	 \mathbf{G}=\operatorname{diag}\left\{e^{2\big(\xi(Z)+\gamma(Y)\big)},e^{2\big(\eta(Z)
	 +\lambda(X)\big)},e^{2\psi(Z)}\right\}.
\end{equation}
The above dislocation distribution corresponds to the following torsion $2$-forms (cf. \eqref{burger})
\begin{equation}
\mathcal{T}^1=b_1(Z)\vartheta^3\wedge\vartheta^1+b_2(X,Y,Z)\vartheta^1\wedge\vartheta^2\,,\quad\mathcal{T}^2=c_1(Z)\vartheta^2\wedge\vartheta^3+c_2(X,Y,Z)\vartheta^1\wedge\vartheta^2\,,\quad\mathcal{T}^3=0\,.
\end{equation}
This represents a distribution of edge dislocations with the following total Burgers' vector density   
\begin{equation}\label{burg-edge}
\begin{split}
\mathbf{b}(X,Y,Z)&=\left(b_1(Z)+b_2(X,Y,Z)\right)\mathbf{e}_1+\left(c_1(Z)+c_2(X,Y,Z)\right)\mathbf{e}_2\\
&=e^{-\xi(Z)-\gamma(Y)}\left[b_1(Z)+b_2(X,Y,Z)\right]\partial_X
+e^{-\eta(Z)-\lambda(X)}\left[c_1(Z)+c_2(X,Y,Z)\right]\partial_Y\,.
\end{split}
\end{equation}
From \eqref{fr-fi}, one obtains
\begin{equation}
\begin{split}
d\vartheta^1=&e^{-\psi(Z)}\xi'(Z)\vartheta^3\wedge\vartheta^1+e^{-\eta(Z)-\lambda(X)}\gamma'(Y)\vartheta^2\wedge\vartheta^1\,,\\ d\vartheta^2=&e^{-\psi(Z)}\eta'(Z)\vartheta^3\wedge\vartheta^2
+e^{-\xi(Z)-\gamma(Y)}\lambda'(X)\vartheta^1\wedge\vartheta^2\,,\qquad d\vartheta^3=0\,.
\end{split}
\end{equation}
Metric compatibility implies the following connection $1$-forms matrix 
\begin{equation}
\label{om-ga}
\boldsymbol{\omega}=[\omega^{\alpha}{}_{\beta}]=
\left(
\begin{array}{ccc}
0&\omega^1{}_2&-\omega^3{}_1\\
-\omega^1{}_2&0&\omega^2{}_3\\
\omega^3{}_1&-\omega^2{}_3&0
\end{array}
\right)\,.
\end{equation}
Cartan's first structural equation gives the following connection $1$-forms 
\begin{equation}
\begin{split}
\omega^1{}_2&=\left(b_2(X,Y,Z)+\gamma'(Y)e^{-\eta(Z)-\lambda(X)}\right)\vartheta^1
+\left(c_2(X,Y,Z)-\lambda'(X)e^{-\xi(Z)-\gamma(Y)}\right)\vartheta^2\,,\\ 
\omega^2{}_3&=\left(c_1(Z)+\eta'(Z)e^{-\psi(Z)}\right)\vartheta^2\,,\qquad
\omega^3{}_1=\left(b_1(Z)-\xi'(Z)e^{-\psi(Z)}\right)\vartheta^1\,.
\end{split}
\end{equation}
The second structural equation, i.e., $\mathcal{R}^{\alpha}{}_{\beta}=0$ is trivially satisfied if one assumes that
\begin{equation}\label{sec-st} 
\begin{split}
\xi'(Z)=b_1(Z)e^{\psi(Z)}\,,\quad \gamma'(Y)&=-b_2(X,Y,Z)e^{\eta(Z)+\lambda(X)}\,,\quad \eta'(Z)=-c_1(Z)e^{\psi(Z)}\,,\\ \lambda'(X)&=c_2(X,Y,Z)e^{\xi(Z)+\gamma(Y)}\,.
\end{split}
\end{equation}
Thus
\begin{equation}
	\eta'(Z) =-\frac{1}{b_2}\frac{\partial b_2}{\partial Z},~~\lambda'(X)
	=-\frac{1}{b_2}\frac{\partial b_2}{\partial X},~~\gamma'(Y)
	=-\frac{1}{c_2}\frac{\partial c_2}{\partial Y},~~ \xi'(Z)
	=-\frac{1}{c_2}\frac{\partial c_2}{\partial Z},~~ \psi(Z)
	=\ln\Big(\frac{-1}{b_1c_2}\frac{\partial c_2}{\partial Z}\Big)\,,
\end{equation}
where one needs to have $\frac{\partial b_2}{\partial Z}=-\frac{c_1 b_2}{b_1 c_2}\frac{\partial c_2}{\partial Z}$ and $-\frac{1}{b_1c_2}\frac{\partial c_2}{\partial Z}>0$. If we assume that $b_2$ and $c_2$ are separable in $X$, $Y$, and $Z$, i.e., $b_2=b_{2X}(X)b_{2Y}(Y)b_{2Z}(Z)$ and $c_2=c_{2X}(X)c_{2Y}(Y)c_{2Z}(Z)$, then
\begin{equation}
e^{\eta(Z)}=\frac{C_1}{b_{2Z}(Z)}\,,\quad e^{\lambda(X)}=\frac{C_2}{b_{2X}(X)}\,,\quad e^{\gamma(Y)}=\frac{C_3}{c_{2Y}(Y)}\,,\quad e^{\xi(Z)}=\frac{C_4}{c_{2Z}(Z)}\,,\quad e^{\psi(Z)}=-\frac{c_{2Z}(Z)'}{b_1(Z)c_{2Z}(Z)}\,,
\end{equation}
where $C_i$, $i=1,\dots,4$ are constants of integration. The compatibility conditions are written as
\begin{equation}
	C_1C_2\,b_{2Y}(Y)=\frac{c'_{2Y}(Y)}{c_{2Y}(Y)}\,,\quad 
	C_3C_4\,c_{2X}(X)=-\frac{b'_{2X}(X)}{b_{2X}(X)}\,,\quad 
	\frac{b'_{2Z}(Z)}{b_{2Z}(Z)}=-\frac{c_1(Z)}{b_1(Z)}\frac{c'_{2Z}(Z)}{c_{2Z}(Z)}\,.
\end{equation}
Therefore, we have the material manifold \eqref{G-edge} for the edge dislocation distributions with the Burgers' vector density \eqref{burg-edge}. For the sake of simplicity of calculations, in the remaining of this section we consider two simplified cases of the distribution \eqref{burg-edge}: (i) $b_2(X,Y,Z)=c_2(X,Y,Z)=0$, $\gamma(Y)=0$, $\lambda(X)=0$, $\psi(Z)=0$, and (ii) $b_2(X,Y,Z)=c_2(X,Y,Z)=0$, $\gamma(Y)=0$, $\lambda(X)=0$, $c_1(Z)=0$, $\eta(Z)=0$.

\paragraph{Case (i).}From \eqref{burg-edge}, the Burgers' vector density reads $\mathbf{b}=\mathbf{b}(Z)=b_1(Z)\mathbf{e}_1+c_1(Z)\mathbf{e}_2=b_1(Z)e^{-\xi(Z)}\partial_X+e^{-\eta(Z)}c_1(Z)\partial_Y$, where, using \eqref{sec-st}, $\xi'(Z)=b_1(Z)$ and $\eta'(Z)=-c_1(Z)$. The material metric \eqref{G-edge} is simplified as $\mathbf{G}=\operatorname{diag}\left\{e^{2\xi(Z)},e^{2\eta(Z)},1\right\}$. Looking for solutions of the form $(x,y,z)=(X,Y,\alpha Z)$, the incompressibility constraint implies that $J=\frac{\alpha}{e^{\xi(Z)+\eta(Z)}}=1$, and thus, $\xi(Z)+\eta(Z)=\ln\alpha$. This means that $\xi'(Z)+\eta'(Z)=0$, and hence, $c_1(Z)=b_1(Z)$. Choosing orthonormal vectors $\mathbf{N}_1=e^{-\xi(Z)}\partial_X$,  $\mathbf{N}_2=e^{-\eta(Z)}\partial_Y$, and $\mathbf{N}_3=\partial_Z$ as the orthotropic axes, the invariants of the energy function are obtained from \eqref{invaror} as follows
\begin{equation}
I_1=\alpha^2+e^{-2\xi(Z)}+\frac{1}{\alpha^2}e^{2\xi(Z)}\,,\quad I_2=\frac{1}{\alpha^2}+e^{2\xi(Z)}+\alpha^2e^{-2\xi(Z)}\,,\quad I_5=I_4^2=e^{-4\xi(Z)}\,,\quad I_7=I_6^2=\frac{e^{4\xi(Z)}}{\alpha^4}\,.
\end{equation}
Therefore, the non-zero components of the Cauchy stress tensor read
\begin{align}
\hat{\sigma}^{xx}&=2e^{-2\xi(Z)}\left[W_{I_1}+\Big(\frac{e^{2\xi(Z)}}{\alpha^2}+\alpha^2\Big)W_{I_2}+W_{I_4}\right]+4e^{-4\xi(Z)}W_{I_5}-p(Z)\,,\\
\hat{\sigma}^{yy}&=\frac{2}{\alpha^2}e^{2\xi(Z)}\left[W_{I_1}+\Big(e^{-2\xi(Z)}+\alpha^2\Big)W_{I_2}+W_{I_6}\right]+\frac{4}{\alpha^4}e^{4\xi(Z)}W_{I_7}-p(Z)\,,\\
\hat{\sigma}^{zz}&=2\alpha^2\left[W_{I_1}+\Big(e^{-2\xi(Z)}+\frac{e^{2\xi(Z)}}{\alpha^2}\Big)W_{I_2}\right]-p(Z)\,.
\end{align}
Equilibrium equations imply that $\hat{\sigma}^{zz}=C$, where $C$ is a constant. Vanishing of the traction vector on surfaces parallel to the $\mathrm{X}-\mathrm{Y}$ plane gives the pressure as
\begin{equation}
P(Z)=2\alpha^2\left[W_{I_1}+\Big(e^{-2\xi(Z)}+\frac{e^{2\xi(Z)}}{\alpha^2}\Big)W_{I_2}\right]\,.
\end{equation}

\paragraph{Case (ii).} The Burgers' vector density is given by $\mathbf{b}=\mathbf{b}(Z)=b_1(Z)\mathbf{e}_1=b_1(Z)e^{-\xi(Z)}\partial_X$, where $\xi'(Z)=b_1(Z)$. From \eqref{G-edge}, the material metric is $\mathbf{G}=\operatorname{diag}\left\{e^{2\xi(Z)},1,e^{2\psi(Z)}\right\}$. We then look for solutions of the form $(x,y,z)=(X,Y,\alpha Z)$. Incompressibility implies that $J=\frac{\alpha}{e^{\xi(Z)+\psi(Z)}}=1$, and hence, $\xi(Z)+\psi(Z)=\ln\alpha$. The orthotropic axes are $\mathbf{N}_1=e^{-\xi(Z)}\partial_X$, $\mathbf{N}_2=\partial_Y$, and $\mathbf{N}_3=e^{-\psi(Z)}\partial_Z$. The invariants of the strain energy function read
\begin{equation}
I_1=1+e^{-2\xi(Z)}+e^{2\xi(Z)}\,,\quad I_2=1+e^{2\xi(Z)}+e^{-2\xi(Z)}\,,\quad I_5=I_4^2=e^{-4\xi(Z)}\,,\quad I_6=I_7=1\,.
\end{equation}
The non-zero components of the Cauchy stress are given as
\begin{align}
\hat{\sigma}^{xx}&=2W_{I_2}+2e^{-2\xi(Z)}\left(W_{I_1}+W_{I_2}+W_{I_4}\right)+4e^{-4\xi(Z)}W_{I_5}-p(Z)\,,\\
\hat{\sigma}^{yy}&=2W_{I_1}+2W_{I_2}\left(e^{-2\xi(Z)}+e^{2\xi(Z)}\right)+2W_{I_6}+4W_{I_7}-p(Z)\,,\\
\hat{\sigma}^{zz}&=2\left(W_{I_2}+e^{2\xi(Z)}(W_{I_1}+W_{I_2})\right)-p(Z)\,.
\end{align}
The equilibrium equation and the vanishing of traction vector on surfaces parallel to $\mathrm{X}-\mathrm{Y}$ plane yield
\begin{equation}
p(Z)=2\left(W_{I_2}+e^{2\xi(Z)}(W_{I_1}+W_{I_2})\right)\,.
\end{equation} 

\subsection{A Spherically-Symmetric Distribution of Point Defects in a Transversely Isotropic Ball}
\label{eballc}

\begin{figure}
	\centering
	\begin{center}
		\makebox[0pt]{\includegraphics[trim = 14cm 0.4cm 14cm 0.5cm, clip, scale=0.35]{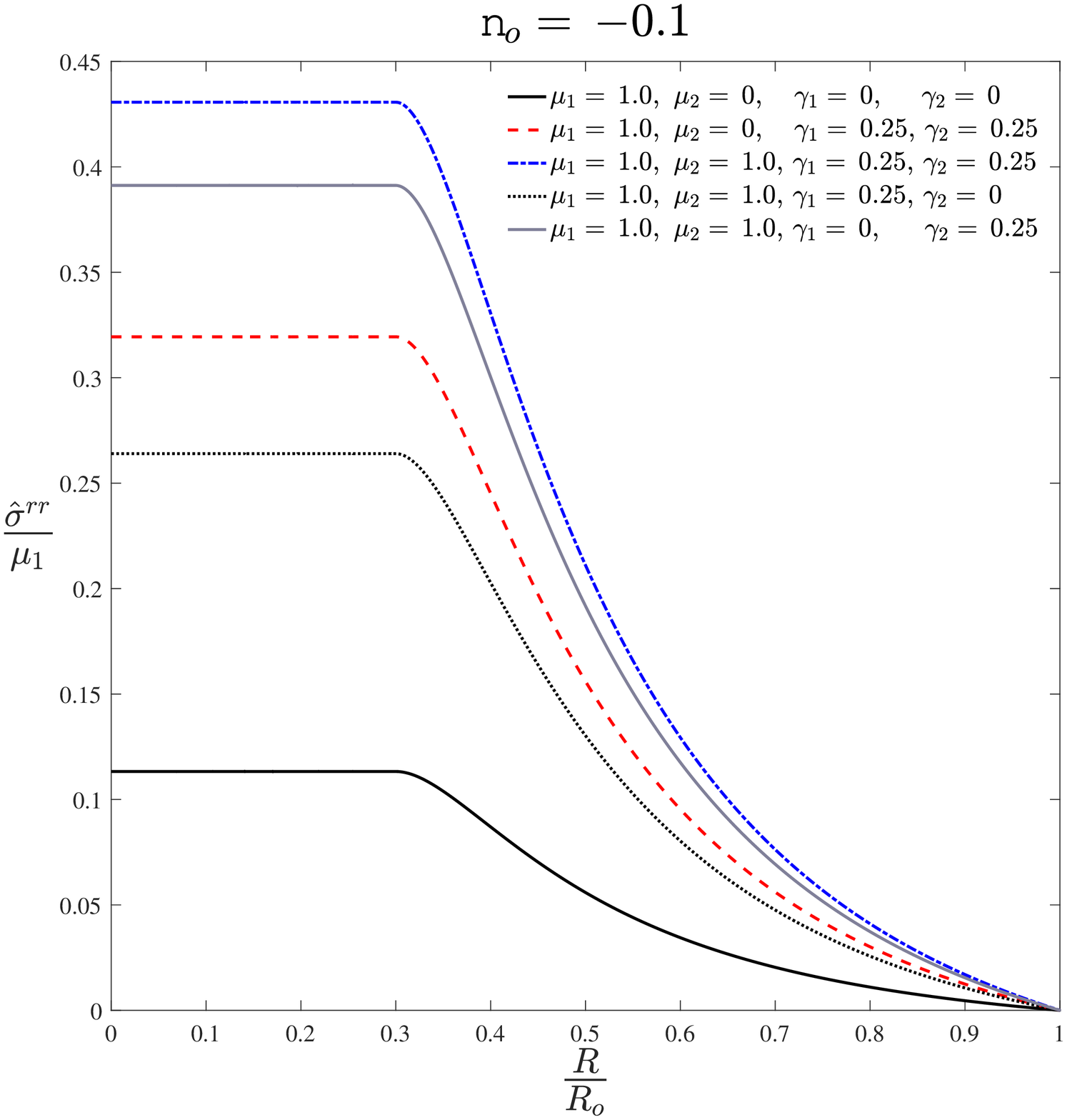}
			\includegraphics[trim = 14cm 0.4cm 14cm 0.5cm, clip, scale=0.35]{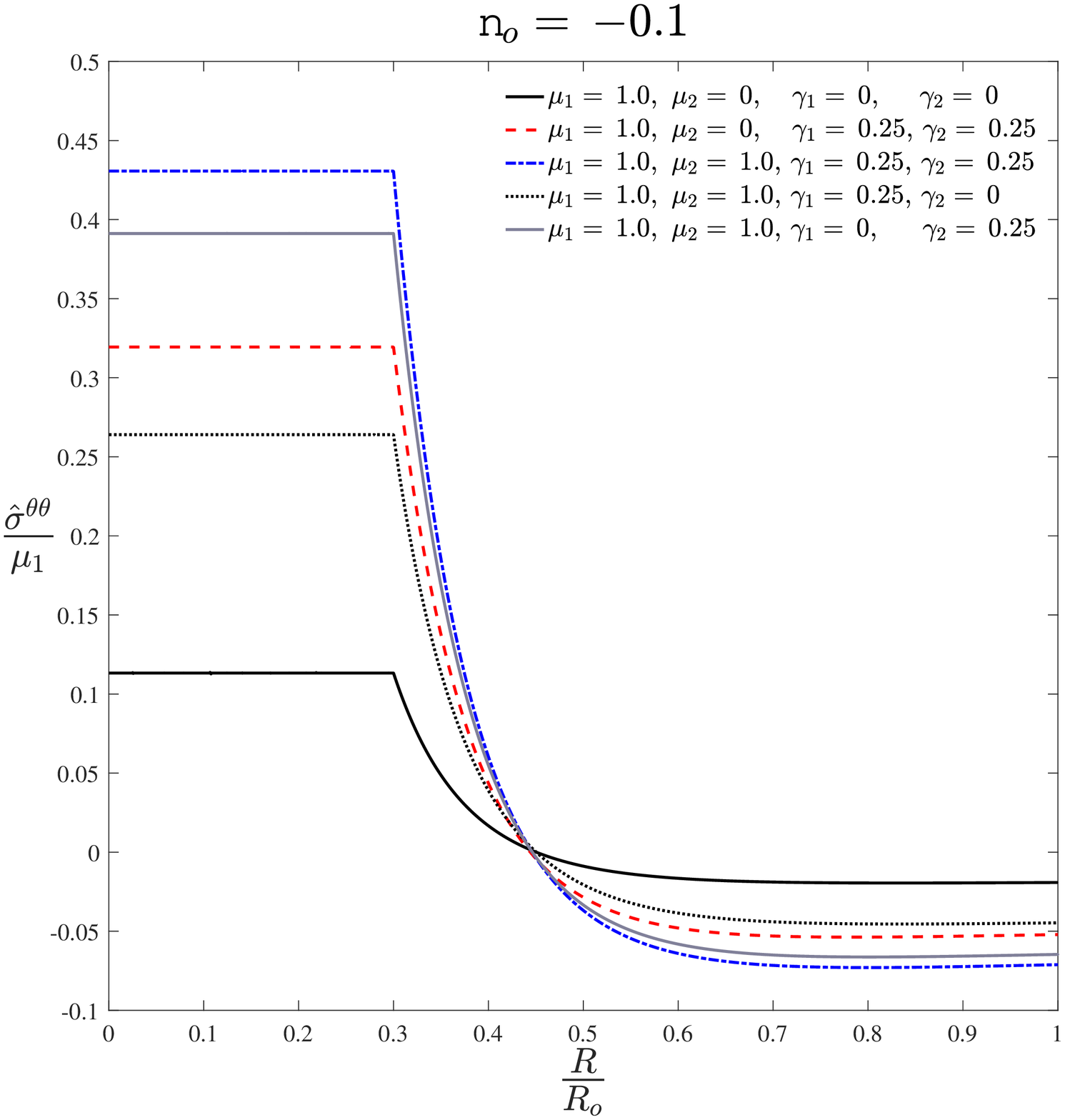}}
	\end{center}
	\vspace*{-0.15in}
	\caption{$\hat{\sigma}^{rr}$ and $\hat{\sigma}^{\theta\theta}$ distribution for different values of the constitutive parameters for the point defect distribution \eqref{po-uni} with $R_i/R_o=0.3$ and $\mathsf{n}_o=-0.1$.} 
	\label{si-rt-po}
\end{figure}

In this section, we calculate the stress field of a spherically-symmetric distribution of point defects in a transversely isotropic ball of radius $R_o$. The material manifold of a medium with distributed point defects is a flat Weyl manifold \citep{YavariGoriely2012c}. Let us assume that the material preferred direction is radial, i.e., $\mathbf{N}=\hat{\mathbf R}$,\footnote{Note that $\hat{\mathbf{R}}=\frac{1}{f(R)}\mathbf{E}_R$ is the unit vector identifying the material preferred direction, where $\mathbf{E}_R=\frac{\partial}{\partial R}$ such that $\langle\!\langle\mathbf{E}_R,\mathbf{E}_R\rangle\!\rangle_{\mathbf{G}}=G_{RR}$.} where $\hat{\mathbf R}$ is a unit vector in the radial direction. The material metric for the body with a radial distribution of point defects in the spherical coordinates $\left(R,\Theta,\Phi\right)$ reads $\mathbf{G}=\operatorname{diag}\left\{f^2(R),R^2,R^2\sin^2\Theta\right\}$, where
\begin{equation}\label{point}
	f(R)=\frac{1-\mathsf{n}(R)}{1-\frac{1}{R^3}\int_{0}^{R}3y^2\mathsf{n}(y)dy}\,.
\end{equation}
We endow the ambient space with the flat Euclidean metric $\mathbf{g}=\operatorname{diag}\left\{1,r^2,r^2\sin^2\theta\right\}$ in the spherical coordinates $\left(r,\theta,\phi\right)$. Given an embedding of the form $\left(r,\theta,\phi\right)=\left(r\left(R\right),\Theta,\Phi\right)$, the deformation gradient is written as $\mathbf{F}=\operatorname{diag}\{r'(R),1,1\}$. The right Cauchy-Green deformation tensor reads $\mathbf{C}=\operatorname{diag}\left\{\frac{{r'}^2(R)}{f^2(R)},\frac{r^2(R)}{R^2},\frac{r^2(R)}{R^2}\right\}$. Assuming incompressibility, the Jacobean is expressed as
\begin{equation}\label{jacob-sph}
J=\sqrt{\frac{\mathrm{det} \mathbf g}{\mathrm{det} \mathbf G}} \mathrm{det} \mathbf F=\frac{r^2(R)r'(R)}{R^2f(R)}=1\,.
\end{equation}
This gives $r(R)=\left(\int_{0}^{R}3\xi^2f(\xi)d\xi\right)^{\frac{1}{3}}$. Using \eqref{invartr}, the invariants are written as
\begin{equation}
\begin{aligned}
	I_1=\operatorname{tr}(\mathbf{C})=\frac{R^4}{r^4(R)}+2\frac{r^2(R)}{R^2}\,, \quad
	I_2=\frac{1}{2}(\operatorname{tr}(\mathbf{C}^2)-\operatorname{tr}(\mathbf{C})^2)
	=\frac{r^4(R)}{R^4}+\frac{2R^2}{r(R)^2}\,,\quad I_5=I_4^2=\frac{R^8}{ r^8(R)}\,.
\end{aligned}
\end{equation}
Using \eqref{jacob-sph}, the non-zero stress components read\footnote{When the body is defect-free, $f(R)=1$, and thus, $I_1=I_2=3$ and $I_4=I_5=1$. If one assumes that the stress vanishes in this case, one has (see  \citep{merodio2003instabilities,vergori2013anisotropic} for similar conditions)
	\begin{equation}
	\left(2 
	W_{I_5}+ W_{I_4}\right)|_{I_1=I_2=3, I_4=I_5=1}=0\,.
	\label{comptr}
	\end{equation}
}
\begin{equation}\label{st-po-g}
\begin{aligned}
	\hat{\sigma}^{rr}&=2\frac{R^4}{r^4(R)}\left(W_{I_1}+W_{I_4}\right)
	+4\frac{R^2}{r^2(R)}W_{I_2}+4\frac{R^8}{r^8(R)}W_{I_5}-p(R)\,,\\ 
	\hat{\sigma}^{\theta\theta}&=\hat{\sigma}^{\phi\phi}=2\frac{r^2(R)}{R^2}W_{I_1}
	+2\frac{R^2}{r^2(R)}W_{I_2}+2\frac{r^4(R)}{R^4}W_{I_2}-p(R)\,.
\end{aligned}
\end{equation}
The non-trivial equilibrium equation is simplified to read $\frac{1}{r'(R)}\sigma^{rr}{}_{,R}+\frac{2}{r}\sigma^{rr}-2r\sigma^{\theta\theta}=0$. This gives $p'(R)=q(R)$, where
\begin{equation}\label{pr-poi}
\begin{aligned}
	q(R)=&-\frac{4}{R^3 r^{19}} \Big[f \Big\{R^9 r^{12} \left(W_{I_1}+4 W_{I_2I_2}-4W_{I_2I_5}+W_{I_4}\right)
	+R^3 r^{18} \left(W_{I_1}-4W_{I_2I_2}\right)\\&+R^7 r^{14} \left[W_{I_2}
	-2\left(W_{I_1I_1}+W_{I_1I_4}\right)\right]+2 R^{13} r^8\left(W_{I_1I_1}+2W_{I_1I_4}+W_{I_4I_4}
	+3W_{I_5}\right)\\&+2 R^{11} r^{10}\left(3 W_{I_1I_2}-2 W_{I_1I_5}+3W_{I_2I_4}\right)
	-2 R^5 r^{16}\left(3W_{I_1I_2}+W_{I_2I_4}\right)+8 R^{17} r^4\left(W_{I_1I_5}+W_{I_4I_5}\right)\\
	&+R W_{I_2} r^{20}+12 R^{15}W_{I_2I_5} r^6+8 R^{21}W_{I_5I_5}\Big\}-2 r^3 
	\Big\{R^6 r^{12} \left(W_{I_1}+2W_{I_2I_2}-2W_{I_2I_5}+W_{I_4}\right)\\&+R^4r^{14}
	\left(W_{I_2}-W_{I_1I_1}-W_{I_1I_4}\right)+R^{10} r^8\left(W_{I_1I_1}+2W_{I_1I_4}+W_{I_4I_4}
	+4W_{I_5}\right)\\&+R^8 r^{10} \left(3W_{I_1I_2}-2 W_{I_1I_5}+3W_{I_2I_4}\right)-R^2 r^{16} 
	\left(3W_{I_1I_2}+W_{I_2I_4}\right)+4R^{14} r^4\left(W_{I_1I_5}+W_{I_4I_5}\right)\\
	&-2 W_{I_2I_2} r^{18}+6 R^{12}W_{I_2I_5} r^6+4 R^{18}W_{I_5I_5}\Big\}\Big]\,.
\end{aligned}
\end{equation}
Next, we assume an energy function corresponding to a radially reinforced Mooney-Rivlin spherical ball of the following form
\begin{equation}
W(I_1,I_2,I_4,I_5) =\frac{\mu_1}{2}\left(I_1-3\right)+\frac{\mu_2}{2}\left(I_2-3\right)+\frac{\gamma_1}{2}\left(I_4-1\right)^2
+\frac{\gamma_2}{2}\left(I_5-1\right)^2\,,
\label{sten2}
\end{equation} 
where $\mu_1$ and $\mu_2$ are constants of the Mooney-Rivlin base material, while $\gamma_1$ and $\gamma_2$ are non-negative material constants pertaining to the reinforcement strength in the radial direction. Thus, \eqref{pr-poi} is simplified to read
\begin{equation}
\begin{split}
p'(R)=&-\frac{2}{R^2
	r^{19}} \Big[f \left\{6 R^{12} (\gamma_1-2 \gamma_2) r^8+R^8 (\mu_1-2 \gamma_1) r^{12}+\mu_2
R^6 r^{14}+\mu_1 R^2
r^{18}+\mu_2 r^{20}+28
\gamma_2 R^{20}\right\}\\&-2 R^3 r^3
\left(4 R^6 (\gamma_1-2 \gamma_2) r^8+R^2 (\mu_1-2 \gamma_1) r^{12}+\mu_2 r^{14}+16
\gamma_2 R^{14}\right)\Big]\,.
\end{split}
\end{equation}
The stress components are also simplified and read
\begin{equation}\label{st-po}
\begin{aligned}
	\hat{\sigma}^{rr}&=\frac{R^4}{r^4}\Big[{\mu_1}+2\gamma_1\Big(\frac{R^4}{r^4}-1\Big)\Big]
	+2{\mu_2}\frac{R^2}{r^2}+4\gamma_2\frac{R^8}{r^8}\Big(\frac{R^8}{r^8}-1\Big)-p\,,\\ \hat{\sigma}^{\theta\theta}	
	&=\hat{\sigma}^{\phi\phi}={\mu_1}\frac{r^2}{R^2}+{\mu_2}\frac{R^2}{r^2}+{\mu_2}\frac{r^4}{R^4}-p\,.
\end{aligned}
\end{equation}
Assuming that the boundary of the ball is traction-free, one obtains
\begin{equation}
	p(R_o)=\frac{R^4_o}{r^4(R_o)})\Big[{\mu_1}+2\gamma_1\Big(\frac{R^4_o}{r^4(R_o)}-1\Big)\Big]
	+2{\mu_2}\frac{R^2_o}{r^2(R_o)}+4\gamma_2\frac{R^8_o}{r^8(R_o)}\Big(\frac{R^8_o}{r^8(R_o)}-1\Big)\,.
\end{equation}
Thus
\begin{equation}
\begin{split}
	p(R)=&2\int_{R}^{R_o}\frac{1}{\xi^2
	r^{19}(\xi)} \Big[f(\xi) \Big\{6 \xi^{12} (\gamma_1-2 \gamma_2) r^8(\xi)
	+\xi^8 (\mu_1-2 \gamma_1) r^{12}(\xi)+\mu_2\xi^6 r^{14}(\xi)+\mu_1 \xi^2r^{18}(\xi)\\&+\mu_2 r^{20}
	+28\gamma_2 \xi^{20}\Big\}-2 \xi^3 r^3(\xi)\Big(4 \xi^6 (\gamma_1-2 \gamma_2) r^8(\xi)
	+\xi^2 (\mu_1-2 \gamma_1) r^{12}+\mu_2 r^{14}(\xi)\\&+16\gamma_2 \xi^{14}\Big)\Big]d\xi+p(R_o)\,.
\end{split}
\end{equation}
Let us consider the following distribution of point defects in the ball
\begin{equation}\label{po-uni}
\mathsf{n}(R)=
\begin{cases}\mathsf{n}_o~~\quad 0\leq R\leq R_i\,,\\
0~\quad ~ R_i<R\leq R_o\,.
\end{cases}
\end{equation}
Therefore, from \eqref{point}
\begin{equation}\label{f-po}
f(R)=
\begin{cases}
1\,,\quad\quad\quad\quad\quad\quad\quad \quad~~ 0\leq R\leq R_i\,,\\
(1-\mathsf{n}_o(R_i/R)^3)^{-1}\,,~ R_i<R\leq R_o\,,
\end{cases}
\end{equation}
and hence
\begin{equation}\label{r-po}
r(R)=
\begin{cases}
R\,,\quad ~~~~~~~~~~~~~~~~~~~~~~~~~~~~~~~~~~~0\leq R\leq R_i\,,\\
\left[R^3+\mathsf{n}_oR_i^3\ln\frac{(R/R_i)^3-\mathsf{n}_o}{1-\mathsf{n}_o}\right]^{1/3}\,,~ R_i<R\leq R_o\,.
\end{cases}
\end{equation}
Fig.~\ref{si-rt-po} shows the stress field variation for the point defect distribution \eqref{po-uni}, where $R_i/R_o=0.3$ and $\mathsf{n}_o=-0.1$ for different values of the reinforcement and base material constants in \eqref{sten2}.

\begin{remark}
Consider an arbitrary nonlinear incompressible transversely isotropic spherical ball of radius $R_o$ such that the material preferred direction is radial. Suppose that the ball is subject to a uniform pressure on its boundary and has the point defect distribution \eqref{po-uni}. Then, in the ball $R\leq R_i$, the stress is uniform and hydrostatic. Interestingly, the value of the hydrostatic stress inside the ball $R\leq R_i$ has an explicit dependence on the reinforcement parameters (see Fig.~\ref{si-rt-po}). To show this, for $R\leq R_i$, $f(R)=1$ and $r(R)=R$, following \eqref{f-po} and \eqref{r-po}, respectively. Therefore, after some simplification, \eqref{pr-poi} implies that $p'(R)=q(R)=\frac{4}{R}\left(W_{I_4}+2W_{I_5}\right)|_{I_1=I_2=3,I_4=I_5=1}=0$, where we used the relation \eqref{comptr}. Hence, for $R\leq R_i$, $p(R)=C$, where $C$ is a constant depending on the reinforcement and base material parameters. From \eqref{st-po-g}, $\hat{\sigma}^{rr}=\hat{\sigma}^{\theta\theta}=\hat{\sigma}^{\phi\phi}=2\left(W_{I_1}+2W_{I_2}\right)|_{I_1=I_2=3,I_4=I_5=1}-C$ for $R\leq R_i$.
\end{remark}

\section{Concluding Remarks}
\label{conc}

Despite the crucial role that anisotropy plays in the overall response of materials in the presence of large strains, the study of defects in nonlinear solids has been overwhelmingly restricted to isotropic materials to this date. In this paper, we presented a few analytical solutions for the stress fields induced by distributed line and point defects in nonlinear anisotropic solids. We considered a parallel cylindrically-symmetric distribution of screw dislocations in infinite orthotropic and monoclinic media, and also, a cylindrically-symmetric distribution of parallel wedge disclinations in an orthotropic medium. 
Because the material manifold is endowed with a nontrivial Riemannian metric that explicitly depends on the defect distribution, the material preferred directions, and hence, the class of anisotropy of the defective body are, in general, different in the reference and current configurations. We observed, in particular, that for a cylindrically-symmetric distribution of screw dislocations, assuming that the body is orthotropic in the reference (current) configuration, it is monoclinic in its current (reference) configuration. 
We found that for an arbitrary cylindrically-symmetric distribution of parallel screw dislocations and a uniform wedge disclination distribution, the stress field is logarithmically singular on the dislocation and disclination axes, respectively, unless the axial deformation is suppressed. These stress singularities are inherent to the anisotropic effects due to the radial fiber-reinforcement, and do not, in particular, arise in isotropic materials. This observation demonstrates the significance of taking material anisotropy into consideration in the analysis of solids with distributed defects.
For a single screw dislocation, we employed the standard reinforcing model and discussed the conditions that guarantee that the energy per unit length and the resultant axial force are finite for a fiber-reinforced material as long as the isotropic base material has a finite energy per unit length and a finite axial force. For a distribution of edge dislocations the resulting stress are calculated when the medium is orthotropic. Finally, we studied a spherically-symmetric distribution of point defects in a transversely isotropic spherical ball. We showed that for an arbitrary incompressible transversely isotropic ball with the radial material preferred direction, a uniform point defect distribution results in a uniform hydrostatic stress field inside the spherical region the distribution is supported in.     
The role that anisotropy plays in the dynamics, stability, and interactions of defects at finite strains are exciting problems that will be the subjects of future communications.  

\section*{Acknowledgement}

This work was partially supported by NSF -- Grant No. CMMI 1561578,  ARO Grant No. W911NF-18-1-0003, and AFOSR -- Grant No. FA9550-12-1-0290.  

\bibliographystyle{abbrvnat}
\bibliography{biblio}

\begin{thebibliography}{76}
\providecommand{\natexlab}[1]{#1}
\providecommand{\url}[1]{\texttt{#1}}
\expandafter\ifx\csname urlstyle\endcsname\relax
  \providecommand{\doi}[1]{doi: #1}\else
  \providecommand{\doi}{doi: \begingroup \urlstyle{rm}\Url}\fi

\bibitem[Acharya(2001)]{acharya2001model}
A.~Acharya.
\newblock A model of crystal plasticity based on the theory of continuously
  distributed dislocations.
\newblock \emph{Journal of the Mechanics and Physics of Solids}, 49\penalty0
  (4):\penalty0 761--784, 2001.

\bibitem[Amar and Goriely(2005)]{amar2005growth}
M.~B. Amar and A.~Goriely.
\newblock Growth and instability in elastic tissues.
\newblock \emph{Journal of the Mechanics and Physics of Solids}, 53\penalty0
  (10):\penalty0 2284--2319, 2005.

\bibitem[Bilby et~al.(1955)Bilby, Bullough, and Smith]{bilby1955continuous}
B.~Bilby, R.~Bullough, and E.~Smith.
\newblock Continuous distributions of dislocations: a new application of the
  methods of non-riemannian geometry.
\newblock In \emph{Proceedings of the Royal Society of London A: Mathematical,
  Physical and Engineering Sciences}, volume 231, pages 263--273. The Royal
  Society, 1955.

\bibitem[Brake(2012)]{brake2012analytical}
M.~Brake.
\newblock An analytical elastic-perfectly plastic contact model.
\newblock \emph{International Journal of Solids and Structures}, 49\penalty0
  (22):\penalty0 3129--3141, 2012.

\bibitem[Brake(2015)]{brake2015analytical}
M.~Brake.
\newblock An analytical elastic plastic contact model with strain hardening and
  frictional effects for normal and oblique impacts.
\newblock \emph{International Journal of Solids and Structures}, 62:\penalty0
  104--123, 2015.

\bibitem[Clayton(2017)]{clayton2017finsler}
J.~Clayton.
\newblock Finsler geometry of nonlinear elastic solids with internal structure.
\newblock \emph{Journal of Geometry and Physics}, 112:\penalty0 118--146, 2017.

\bibitem[Demirkoparan and Merodio(2017)]{demirkoparan}
H.~Demirkoparan and J.~Merodio.
\newblock Bulging bifurcation of inflated circular cylinders of doubly
  fiber-reinforced hyperelastic material under axial loading and swelling.
\newblock \emph{Mathematics and Mechanics of Solids}, 22\penalty0 (4):\penalty0
  666--682, 2017.

\bibitem[Derezin and Zubov(2011)]{derezin2011discl}
S.~V. Derezin and L.~M. Zubov.
\newblock Disclinations in nonlinear elasticity.
\newblock \emph{Zeitschrift f{\"u}r Angewandte Mathematik und Mechanik (ZAMM)},
  91\penalty0 (6):\penalty0 433--442, 2011.

\bibitem[do~Carmo(1992)]{Docarmo1992}
M.~do~Carmo.
\newblock \emph{Riemannian Geometry}.
\newblock Mathematics: Theory \& Applications. Birkh{\"a}user Boston, 1992.
\newblock ISBN 1584883553.

\bibitem[Doyle and Ericksen(1956)]{doyle1956nonlinear}
T.~Doyle and J.~Ericksen.
\newblock Nonlinear elasticity.
\newblock \emph{Advances in Applied Mechanics}, 4:\penalty0 53--115, 1956.

\bibitem[Ericksen(1998)]{ericksen1998non}
J.~Ericksen.
\newblock On nonlinear elasticity theory for crystal defects.
\newblock \emph{International Journal of Plasticity}, 14\penalty0
  (1-3):\penalty0 9--24, 1998.

\bibitem[Eshelby(1949)]{eshelby1949}
J.~Eshelby.
\newblock {LXXXII}. {E}dge dislocations in anisotropic materials.
\newblock \emph{The London, Edinburgh, and Dublin Philosophical Magazine and
  Journal of Science}, 40\penalty0 (308):\penalty0 903--912, 1949.

\bibitem[Fedelich(2004)]{fedelich2004glide}
B.~Fedelich.
\newblock The glide force on a dislocation in finite elasticity.
\newblock \emph{Journal of the Mechanics and Physics of Solids}, 52\penalty0
  (1):\penalty0 215--247, 2004.

\bibitem[Gairola(1979)]{Gairola}
B.~Gairola.
\newblock \emph{Nonlinear elastic problems}.
\newblock F.R.N. Nabarro (Ed.), Dislocations in Solids. North-Holland
  Publishing Co., Amsterdam, 1979.

\bibitem[Ghaednia and Marghitu(2016)]{ghaednia2016permanent}
H.~Ghaednia and D.~B. Marghitu.
\newblock Permanent deformation during the oblique impact with friction.
\newblock \emph{Archive of Applied Mechanics}, 86\penalty0 (1-2):\penalty0
  121--134, 2016.

\bibitem[Giordano et~al.(2009)Giordano, Palla, and Colombo]{giordano2009}
S.~Giordano, P.~Palla, and L.~Colombo.
\newblock Nonlinear elasticity of composite materials.
\newblock \emph{The European Physical Journal B}, 68\penalty0 (1):\penalty0
  89--101, 2009.

\bibitem[Golgoon and Yavari(2017{\natexlab{a}})]{Golgoon2017}
A.~Golgoon and A.~Yavari.
\newblock Nonlinear elastic inclusions in anisotropic solids.
\newblock \emph{Journal of Elasticity}, May 2017{\natexlab{a}}.
\newblock \doi{10.1007/s10659-017-9639-0}.

\bibitem[Golgoon and Yavari(2017{\natexlab{b}})]{golgoon2016torus}
A.~Golgoon and A.~Yavari.
\newblock On the stress field of a nonlinear elastic solid torus with a
  toroidal inclusion.
\newblock \emph{Journal of Elasticity}, 128\penalty0 (1):\penalty0 115--145,
  2017{\natexlab{b}}.

\bibitem[Golgoon et~al.(2016)Golgoon, Sadik, and Yavari]{golgoon2016}
A.~Golgoon, S.~Sadik, and A.~Yavari.
\newblock Circumferentially-symmetric finite eigenstrains in incompressible
  isotropic nonlinear elastic wedges.
\newblock \emph{International Journal of Non-Linear Mechanics}, 84:\penalty0
  116--129, 2016.

\bibitem[Goriely et~al.(2010)Goriely, Moulton, and
  Vandiver]{goriely2010elastic}
A.~Goriely, D.~E. Moulton, and R.~Vandiver.
\newblock Elastic cavitation, tube hollowing, and differential growth in plants
  and biological tissues.
\newblock \emph{Europhysics Letters}, 91\penalty0 (1):\penalty0 18001, 2010.

\bibitem[Head(1967)]{head1967unstable}
A.~Head.
\newblock Unstable dislocations in anisotropic crystals.
\newblock \emph{Physica Status Solidi (b)}, 19\penalty0 (1):\penalty0 185--192,
  1967.

\bibitem[Hooshmand et~al.(2017)Hooshmand, Mills, and
  Ghazisaeidi]{hooshmand2017atomistic}
M.~Hooshmand, M.~Mills, and M.~Ghazisaeidi.
\newblock Atomistic modeling of dislocation interactions with twin boundaries
  in ti.
\newblock \emph{Modelling and Simulation in Materials Science and Engineering},
  25\penalty0 (4):\penalty0 045003, 2017.

\bibitem[Jackson and Green(2005)]{jackson2005finite}
R.~L. Jackson and I.~Green.
\newblock A finite element study of elasto-plastic hemispherical contact
  against a rigid flat.
\newblock \emph{Transactions of the ASME-F-Journal of Tribology}, 127\penalty0
  (2):\penalty0 343--354, 2005.

\bibitem[Jackson et~al.(2013)Jackson, Ghaednia, Lee, Rostami, and
  Wang]{jackson2013contact}
R.~L. Jackson, H.~Ghaednia, H.~Lee, A.~Rostami, and X.~Wang.
\newblock Contact mechanics.
\newblock In \emph{Tribology for Scientists and Engineers}, pages 93--140.
  Springer, 2013.

\bibitem[Jackson et~al.(2015)Jackson, Ghaednia, and Pope]{jackson2015solution}
R.~L. Jackson, H.~Ghaednia, and S.~Pope.
\newblock A solution of rigid--perfectly plastic deep spherical indentation
  based on slip-line theory.
\newblock \emph{Tribology Letters}, 58\penalty0 (3):\penalty0 47, 2015.

\bibitem[Kardel et~al.(2017)Kardel, Ghaednia, Carrano, and
  Marghitu]{kardel2017experimental}
K.~Kardel, H.~Ghaednia, A.~L. Carrano, and D.~B. Marghitu.
\newblock Experimental and theoretical modeling of behavior of 3d-printed
  polymers under collision with a rigid rod.
\newblock \emph{Additive Manufacturing}, 14:\penalty0 87--94, 2017.

\bibitem[Katanaev(2005)]{katanaev2005}
M.~Katanaev.
\newblock Introduction to the geometric theory of defects.
\newblock \emph{arXiv preprint cond-mat/0502123}, 2005.

\bibitem[Kinoshita and Mura(1971)]{kinoshita1971}
N.~Kinoshita and T.~Mura.
\newblock Elastic fields of inclusions in anisotropic media.
\newblock \emph{Physica Status Solidi (a)}, 5\penalty0 (3):\penalty0 759--768,
  1971.

\bibitem[Knowles(1977)]{knowles1977finite}
J.~K. Knowles.
\newblock The finite anti-plane shear field near the tip of a crack for a class
  of incompressible elastic solids.
\newblock \emph{International Journal of Fracture}, 13\penalty0 (5):\penalty0
  611--639, 1977.

\bibitem[Kondo(1955{\natexlab{a}})]{kondo1955geometry}
K.~Kondo.
\newblock Geometry of elastic deformation and incompatibility.
\newblock \emph{Memoirs of the {U}nifying {S}tudy of the {B}asic {P}roblems in
  {E}ngineering {S}cience by {M}eans of {G}eometry}, 1:\penalty0 5--17,
  1955{\natexlab{a}}.

\bibitem[Kondo(1955{\natexlab{b}})]{kondo1955non}
K.~Kondo.
\newblock Non-{R}iemannian geometry of imperfect crystals from a macroscopic
  viewpoint.
\newblock \emph{Memoirs of the {U}nifying {S}tudy of the {B}asic {P}roblems in
  {E}ngineering {S}cience by {M}eans of {G}eometry}, 1:\penalty0 6--17,
  1955{\natexlab{b}}.

\bibitem[Li and Dunn(1998)]{li1998anisotropic}
J.~Y. Li and M.~L. Dunn.
\newblock Anisotropic coupled-field inclusion and inhomogeneity problems.
\newblock \emph{Philosophical Magazine A}, 77\penalty0 (5):\penalty0
  1341--1350, 1998.

\bibitem[Liu et~al.(1982)]{liu1982representations}
I.~Liu et~al.
\newblock On representations of anisotropic invariants.
\newblock \emph{International Journal of Engineering Science}, 20\penalty0
  (10):\penalty0 1099--1109, 1982.

\bibitem[Lothe(1992)]{lothe1992dislocations}
J.~Lothe.
\newblock Dislocations in anisotropic media.
\newblock \emph{Elastic {S}train {F}ields and {D}islocation {M}obility}, 31,
  1992.

\bibitem[Lu and Papadopoulos(2000)]{lu2000covariant}
J.~Lu and P.~Papadopoulos.
\newblock A covariant constitutive description of anisotropic non-linear
  elasticity.
\newblock \emph{Zeitschrift f{\"u}r Angewandte Mathematik und Physik (ZAMP)},
  51\penalty0 (2):\penalty0 204--217, 2000.

\bibitem[Marsden and Hughes(1994)]{MarsdenHughes1994}
J.~E. Marsden and T.~J.~R. Hughes.
\newblock \emph{Mathematical Foundations of Elasticity}.
\newblock Dover Publications, New York, 1994.

\bibitem[Merodio and Ogden(2003)]{merodio2003instabilities}
J.~Merodio and R.~Ogden.
\newblock Instabilities and loss of ellipticity in fiber-reinforced
  compressible non-linearly elastic solids under plane deformation.
\newblock \emph{International Journal of Solids and Structures}, 40\penalty0
  (18):\penalty0 4707--4727, 2003.

\bibitem[Merodio and Ogden(2005)]{merodio2005tensile}
J.~Merodio and R.~Ogden.
\newblock Tensile instabilities and ellipticity in fiber-reinforced
  compressible non-linearly elastic solids.
\newblock \emph{International Journal of Engineering Science}, 43\penalty0
  (8):\penalty0 697--706, 2005.

\bibitem[Merodio and Ogden(2006)]{MERODIO200}
J.~Merodio and R.~Ogden.
\newblock The influence of the invariant ${I}_8$ on the stress--deformation and
  ellipticity characteristics of doubly fiber-reinforced non-linearly elastic
  solids.
\newblock \emph{International Journal of Non-Linear Mechanics}, 41\penalty0
  (4):\penalty0 556--563, 2006.

\bibitem[Moulton and Goriely(2011)]{moulton2011anticavitation}
D.~E. Moulton and A.~Goriely.
\newblock Anticavitation and differential growth in elastic shells.
\newblock \emph{Journal of Elasticity}, 102\penalty0 (2):\penalty0 117--132,
  2011.

\bibitem[Mura(1982)]{mura2012}
T.~Mura.
\newblock \emph{Micromechanics of Defects in Solids}.
\newblock Martinus Nijhoff, 1982.

\bibitem[Ozakin and Yavari(2010)]{ozakin2010geometric}
A.~Ozakin and A.~Yavari.
\newblock A geometric theory of thermal stresses.
\newblock \emph{Journal of Mathematical Physics}, 51\penalty0 (3):\penalty0
  032902, 2010.

\bibitem[Ozakin and Yavari(2014)]{ozakin2014affine}
A.~Ozakin and A.~Yavari.
\newblock Affine development of closed curves in {W}eitzenb{\"o}ck manifolds
  and the {B}urgers vector of dislocation mechanics.
\newblock \emph{Mathematics and Mechanics of Solids}, 19\penalty0 (3):\penalty0
  299--307, 2014.

\bibitem[Pence and Tsai(2005)]{pence2005swelling}
T.~J. Pence and H.~Tsai.
\newblock Swelling-induced microchannel formation in nonlinear elasticity.
\newblock \emph{IMA Journal of Applied Mathematics}, 70\penalty0 (1):\penalty0
  173--189, 2005.

\bibitem[Petersen(2006)]{petersen2006riemannian}
P.~Petersen.
\newblock \emph{Riemannian Geometry}, volume 171.
\newblock Springer Science \& Business Media, 2006.

\bibitem[Rosakis and Rosakis(1988)]{rosakis1988screw}
P.~Rosakis and A.~J. Rosakis.
\newblock The screw dislocation problem in incompressible finite elastostatics:
  a discussion of nonlinear effects.
\newblock \emph{Journal of elasticity}, 20\penalty0 (1):\penalty0 3--40, 1988.

\bibitem[Sadik and Yavari(2016)]{Sadik20160659}
S.~Sadik and A.~Yavari.
\newblock Small-on-large geometric anelasticity.
\newblock \emph{Proceedings of the Royal Society of London A}, 472\penalty0
  (2195), 2016.

\bibitem[Sadik and Yavari(2017)]{sadik2015geometric}
S.~Sadik and A.~Yavari.
\newblock Geometric nonlinear thermoelasticity and the time evolution of
  thermal stresses.
\newblock \emph{Mathematics and Mechanics of Solids}, 22\penalty0 (7):\penalty0
  1546--1587, 2017.

\bibitem[Schaefer and Kronm{\"u}ller(1975)]{schaefer1975elastic}
H.~Schaefer and H.~Kronm{\"u}ller.
\newblock Elastic interaction of point defects in isotropic and anisotropic
  cubic media.
\newblock \emph{Physica Status Solidi (b)}, 67\penalty0 (1):\penalty0 63--74,
  1975.

\bibitem[Sozio and Yavari(2017)]{Sozio201712}
F.~Sozio and A.~Yavari.
\newblock Nonlinear mechanics of surface growth for cylindrical and spherical
  elastic bodies.
\newblock \emph{Journal of the Mechanics and Physics of Solids}, 98:\penalty0
  12 -- 48, 2017.

\bibitem[Spencer(1971)]{spencer1971part}
A.~Spencer.
\newblock Part {III}. {T}heory of invariants.
\newblock \emph{Continuum Physics}, 1:\penalty0 239--353, 1971.

\bibitem[Spencer(1982)]{spencer1982formulation}
A.~Spencer.
\newblock The formulation of constitutive equation for anisotropic solids.
\newblock In \emph{Mechanical Behavior of Anisotropic Solids/Comportment
  M{\'e}chanique des Solides Anisotropes}, pages 3--26. Springer, 1982.

\bibitem[Stojanovic et~al.(1964)Stojanovic, Djuric, and
  Vujosevic]{stojanovic1964finite}
R.~Stojanovic, S.~Djuric, and L.~Vujosevic.
\newblock On finite thermal deformations.
\newblock \emph{Archiwum Mechaniki Stosowanej}, 16:\penalty0 103--108, 1964.

\bibitem[Teodosiu(1982)]{teodosiu1982}
C.~Teodosiu.
\newblock \emph{Elastic Models of Crystal Defects}.
\newblock Springer-Verlag, New York, 1982.

\bibitem[Teutonico(1962)]{teutonico1962uniformly}
L.~Teutonico.
\newblock Uniformly moving dislocations of arbitrary orientation in anisotropic
  media.
\newblock \emph{Physical Review}, 127\penalty0 (2):\penalty0 413, 1962.

\bibitem[Triantafyllidis and Abeyaratne(1983)]{triantafy}
N.~Triantafyllidis and R.~Abeyaratne.
\newblock Instabilities of a finitely deformed fiber-reinforced elastic
  material.
\newblock \emph{Journal of applied mechanics}, 50\penalty0 (1):\penalty0
  149--156, 1983.

\bibitem[Truesdell(1953)]{truesdell1953physical}
C.~Truesdell.
\newblock The physical components of vectors and tensors.
\newblock \emph{Zeitschrift f{\"u}r Angewandte Mathematik und Mechanik (ZAMM)},
  33\penalty0 (10-11):\penalty0 345--356, 1953.

\bibitem[Vergori et~al.(2013)Vergori, Destrade, McGarry, and
  Ogden]{vergori2013anisotropic}
L.~Vergori, M.~Destrade, P.~McGarry, and R.~W. Ogden.
\newblock On anisotropic elasticity and questions concerning its finite element
  implementation.
\newblock \emph{Computational Mechanics}, 52\penalty0 (5):\penalty0 1185--1197,
  2013.

\bibitem[Volterra(1907)]{volterra1907equilibre}
V.~Volterra.
\newblock Sur l'{\'e}quilibre des corps {\'e}lastiques multiplement connexes.
\newblock In \emph{Annales scientifiques de l'{\'E}cole normale
  sup{\'e}rieure}, volume~24, pages 401--517, 1907.

\bibitem[Wang et~al.(2013)Wang, Yadav, Hirth, Tom{\'e}, and
  Beyerlein]{wang2013pure}
J.~Wang, S.~Yadav, J.~Hirth, C.~Tom{\'e}, and I.~Beyerlein.
\newblock Pure-shuffle nucleation of deformation twins in
  hexagonal-close-packed metals.
\newblock \emph{Materials Research Letters}, 1\penalty0 (3):\penalty0 126--132,
  2013.

\bibitem[Willis(1964)]{willis1964anisotropic}
J.~Willis.
\newblock Anisotropic elastic inclusion problems.
\newblock \emph{The Quarterly Journal of Mechanics and Applied Mathematics},
  17\penalty0 (2):\penalty0 157--174, 1964.

\bibitem[Willis(1970)]{willis1970stress}
J.~Willis.
\newblock Stress fields produced by dislocations in anisotropic media.
\newblock \emph{Philosophical Magazine}, 21\penalty0 (173):\penalty0 931--949,
  1970.

\bibitem[Willis(1967)]{willis1967second}
J.~R. Willis.
\newblock Second-order effects of dislocations in anisotropic crystals.
\newblock \emph{International Journal of Engineering Science}, 5\penalty0
  (2):\penalty0 171--190, 1967.

\bibitem[Wolfram~Research(2016)]{mathematicap}
I.~Wolfram~Research.
\newblock \emph{Mathematica}.
\newblock Version 11.0. Wolfram Research, Inc., Champaign, Illinois, 2016.

\bibitem[Yavari(2010)]{Yavari2010}
A.~Yavari.
\newblock A geometric theory of growth mechanics.
\newblock \emph{Journal of Nonlinear Science}, 20:\penalty0 781--830, 2010.

\bibitem[Yavari(2016)]{Yavari2016}
A.~Yavari.
\newblock On the wedge dispiration in an inhomogeneous isotropic nonlinear
  elastic solid.
\newblock \emph{Mechanics Research Communications}, 78:\penalty0 55--59, 2016.

\bibitem[Yavari and Goriely(2012{\natexlab{a}})]{YavariGoriely2012a}
A.~Yavari and A.~Goriely.
\newblock Riemann-{C}artan geometry of nonlinear dislocation mechanics.
\newblock \emph{Archive for Rational Mechanics and Analysis}, 205:\penalty0
  59--118, 2012{\natexlab{a}}.

\bibitem[Yavari and Goriely(2012{\natexlab{b}})]{YavariGoriely2012c}
A.~Yavari and A.~Goriely.
\newblock Weyl geometry and the nonlinear mechanics of distributed point
  defects.
\newblock \emph{Proceedings of the Royal Society A}, 468:\penalty0 3902--3922,
  2012{\natexlab{b}}.

\bibitem[Yavari and Goriely(2013{\natexlab{a}})]{yavari2013nonlinear}
A.~Yavari and A.~Goriely.
\newblock Nonlinear elastic inclusions in isotropic solids.
\newblock \emph{Proceedings of the Royal Society A}, 469\penalty0
  (2160):\penalty0 20130415, 2013{\natexlab{a}}.

\bibitem[Yavari and Goriely(2013{\natexlab{b}})]{yavari2013riemann}
A.~Yavari and A.~Goriely.
\newblock Riemann--{C}artan geometry of nonlinear disclination mechanics.
\newblock \emph{Mathematics and Mechanics of Solids}, 18\penalty0 (1):\penalty0
  91--102, 2013{\natexlab{b}}.

\bibitem[Yavari and Goriely(2014)]{yavari2014geometry}
A.~Yavari and A.~Goriely.
\newblock The geometry of discombinations and its applications to semi-inverse
  problems in anelasticity.
\newblock \emph{Proceedings of the Royal Society A}, 470\penalty0
  (2169):\penalty0 20140403, 2014.

\bibitem[Yavari et~al.(2006)Yavari, Marsden, and Ortiz]{Yavari2006}
A.~Yavari, J.~E. Marsden, and M.~Ortiz.
\newblock On the spatial and material covariant balance laws in elasticity.
\newblock \emph{Journal of Mathematical Physics}, 47:\penalty0 85--112, 2006.

\bibitem[Yu(2001)]{yu2001two}
H.~Yu.
\newblock Two-dimensional elastic defects in orthotropic bicrystals.
\newblock \emph{Journal of the Mechanics and Physics of Solids}, 49\penalty0
  (2):\penalty0 261--287, 2001.

\bibitem[Yu et~al.(1995)Yu, Sanday, Rath, and Chang]{yu1995elastic}
H.~Yu, S.~Sanday, B.~Rath, and C.~Chang.
\newblock Elastic fields due to defects in transversely isotropic bimaterials.
\newblock In \emph{Proceedings of the Royal Society of London A: Mathematical,
  Physical and Engineering Sciences}, volume 449, pages 1--30. The Royal
  Society, 1995.

\bibitem[Zheng and Spencer(1993)]{zheng1993tensors}
Q.-S. Zheng and A.~Spencer.
\newblock Tensors which characterize anisotropies.
\newblock \emph{International Journal of Engineering Science}, 31\penalty0
  (5):\penalty0 679--693, 1993.

\bibitem[Zubov(1997)]{zubov1997}
L.~M. Zubov.
\newblock \emph{Nonlinear Theory of Dislocations and Disclinations in Elastic
  Bodies}, volume~47.
\newblock Springer Science \& Business Media, 1997.

\end{thebibliography}

\end{document}